\documentclass[a4paper,fleqn,usenatbib]{mnras}

\usepackage[T1]{fontenc}
\usepackage{ae,aecompl}

\usepackage{graphicx}	
\usepackage{amsmath}	
\usepackage{amssymb}	
\usepackage{subfigure}
\usepackage{multirow}
\usepackage{xcolor}

\newcommand{\pcc}{\,{\rm cm}^{-3}}

\newcommand{\kel}{\, {\rm K}}

\newcommand{\msun}{\, {\rm M}_\odot}
\newcommand{\nh}{n_{_{\rm H}}}

\newcommand{\pc}{\, {\rm pc}}

\newcommand{\myr}{\, {\rm Myr}}

\newcommand{\ug}{\, {\rm \mu G}}

\newcommand{\kms}{\, {\rm km \, s^{-1}}}

\newcommand{\subbroad}{_{_{\rm >0.2pc}}}

\newcommand{\subCOLL}{_{_{\rm COLL}}}

\newcommand{\subCOUPLE}{_{_{\rm COUPLE}}}

\newcommand{\subDRAG}{_{_{\rm DRAG}}}
\newcommand{\subDYN}{_{_{\rm DYN}}}
\newcommand{\subDYNO}{_{_{\rm DYN:O}}}
\newcommand{\subfact}{_{\rm fact}}

\newcommand{\subH}{_{_{\rm H}}}
\newcommand{\subJEANS}{_{_{\rm JEANS}}}

\newcommand{\subMAG}{_{_{\rm MAG}}}
\newcommand{\subMAX}{_{_{\rm MAX}}}

\newcommand{\subMIN}{_{_{\rm MIN}}}
\newcommand{\submolH}{_{_{\rm H_2}}}
\newcommand{\submolHO}{_{_{\rm H_2:O}}}
\newcommand{\subO}{_{_{\rm O}}}
\newcommand{\subperp}{_{_\perp}}

\newcommand{\subSPH}{_{_{\rm SPH}}}
\newcommand{\subSPINE}{_{_{\rm SPINE}}}

\newcommand{\fwhm}{\mbox{\sc fwhm}_{_\Sigma}}

\newcommand{\subOperp}{_{_{\rm O\perp}}}
\newcommand{\subAD}{_{_{\rm AD}}} 
\newcommand{\subADO}{_{_{\rm AD:O}}}

\title[Magntised filament widths]{The widths of magnetised filaments in molecular clouds}

\author[Priestley \& Whitworth]{
F. D. Priestley\thanks{Email: priestleyf@cardiff.ac.uk} and A. P. Whitworth
\\
School of Physics and Astronomy, Cardiff University, Queen's Buildings, The Parade, Cardiff CF24 3AA, UK \\
}

\date{Accepted XXX. Received YYY; in original form ZZZ}

\pubyear{2022}

\begin{document}
\label{firstpage}
\pagerange{\pageref{firstpage}--\pageref{lastpage}}
\maketitle

\begin{abstract}
Filaments are {an ubiquitous} feature of molecular clouds, and appear to play a critical role in assembling the material to form stars. {{The dominant} filaments are observed to have a rather narrow range of widths around $\sim 0.1 \pc$, and to be preferentially aligned perpendicularly to the direction of the local magnetic field.} We have previously argued that the observed filament widths can be explained if filaments are formed by converging, mildly supersonic flows, resulting from large-scale turbulent motions {in} the parent molecular cloud. Here we demonstrate that the introduction of a magnetic field perpendicular to the filament long axis does not greatly alter this conclusion, {as long as the mass-to-flux ratio is supercritical}. The distribution of widths for {supercritical} magnetised filaments formed via this mechanism is peaked at slightly higher values, and is slightly broader, than for non-magnetised filaments, but still reproduces the basic properties of the width distributions derived from far-infrared observations of molecular clouds. {In contrast, subcritical filaments have width distributions with a fundamentally different shape, and typically have much larger widths than those observed.} Both subcritical and supercritical filaments are consistent with the observed lack of correlation between filament widths and filament surface densities.
\end{abstract}

\section{Introduction}

It appears that the magnetic field in the interstellar medium (ISM) -- or at least the component in the plane of the sky -- transitions from being mostly parallel to the low-density structures, to being mostly perpendicular to the high-density structures \citep{PalmeirimPetal2013,soler2013,soler2017}. This means that the dense filaments hosting most low-mass star formation \citep{andre2010,konyves2015,konyves2020}, and also those involved in high-mass star formation via {hub-and-spoke} systems \citep{williams2018,watkins2019,anderson2021}, are usually {orthogonal} to the large-scale magnetic field \citep{pattle2017,pattle2021,arzoumanian2021,kwon2022}.

{On the basis of maps of far-infrared thermal dust-continuum emission,} \citet{arzoumanian2011,arzoumanian2019} argue that the surface-density profiles of filaments in molecular clouds have a characteristic full-width at half-maximum ($\fwhm$) of $\sim 0.1 \pc$, independent of other filament or parent-cloud properties. {Studies based on molecular line observations, rather than thermal dust emission, do not generally support this `characteristic width'} \citep{panopoulou2014,hacar2018,suri2019,alvarez2021,schmiedeke2021,li2022}, but this {can be} explained by the non-linear relation between line intensity and surface-density, due to chemical and radiative transfer effects \citep{priestley2020}. Although \citet{howard2019,howard2021} show that variations in dust temperature and dust opacity may cause similar issues with continuum data, and although \citet{panopoulou2017} show that the averaging process used by \citet{arzoumanian2011,arzoumanian2019} results in an artificially narrow distribution of $\fwhm$s, nonetheless it seems to be the case that the vast majority of observed filaments have $\fwhm$s in the range $\sim 0.02-0.2 \pc$, and not, for example, $\sim 0.4 \pc$.

{A number of explanations for a universal filament width have been advanced. \citet{FischeraJMartinPG2012} show that if one considers an isothermal filament growing quasistatically (and ignores both the early stages when the surface density is arguably too small for the filament to be detected, and the late stages when the filament is expected to fragment into cores), then the $\fwhm$s are typically in the range $\sim 0.05\pc$ to $\sim 0.15\pc$, for typical values of the external pressure in low-mass star forming regions. \citet{HeitschF2013a} and \citet{PalmeirimPetal2013} have emphasised the importance of accretion onto a filament, both in terms of regulating the growth of the line-density, and in terms of exciting internal turbulence. \citet{HennebellePAndreP2013} argue that, if this turbulence is then dissipated by ion-neutral friction, there is a characteristic inner $\fwhm$ of order $0.1\pc$, which is almost independent of the net surface-density. There are several uncertain terms in their analysis (viz. the scaling of the magnetic field with the density, the efficiency with which the accretion energy is converted into turbulence, the dependence of the accretion rate on Larson's scaling relations and the ion-neutral coupling coefficient; see their Section 2), but none of these is likely to change the final estimates.}

Simulations of turbulent molecular gas show that the filamentary structures formed do indeed appear to have a characteristic $\fwhm$ which is comparable to that observed \citep{kirk2015,federrath2016,priestley2020}. These structures form in regions where the turbulent velocity field converges \citep{smith2016,priestley2020}, and this {finding} has motivated us to develop an idealised model for dynamical filament formation via converging supersonic flows \citep{priestley2022}. For a {flat} distribution of inflow Mach numbers, ${\cal M}$, and {a flat distribution of filament ages,} this naturally results in a peaked distribution of $\fwhm$s --- albeit that, for the full range of ${\cal M}$ considered {($1\lesssim{\cal M}\lesssim5$)}, the peak is at $\sim 0.03\pc$ rather than $\sim 0.1\pc$.

The model developed in \citet{priestley2022} is purely hydrodynamical. The addition of a magnetic field is expected to result in broader filaments, due to the additional magnetic pressure. Furthermore, if the field has a component perpendicular to the filament axis, the cylindrical symmetry of the filament is broken, and the observed $\fwhm$ will depend on the viewing angle. This broadening may move the peak of the $\fwhm$ distribution to higher values in better agreement with the \citet{arzoumanian2011,arzoumanian2019} observations, but {it might also} result in filaments which, on average, {are} much broader than those actually observed. In this paper, we extend our converging-flow model of filament formation to include a magnetic field perpendicular to the filament spine, and explore the consequences for the distribution of filament widths. {We show that a key issue is whether the initial magnetic field is {\it supercritical} (too weak to nullify self-gravity) or {\it subcritical} (strong enough to nullify self-gravity).}

{We stress that our models are distinct from those presented by \citet{TomisakaK2014} and \citet{HanawaTTomisakaK2015}, which deal with filaments that have condensed out of a magnetised medium and relaxed to equilibrium. No equilibria are considered here. The supercritical filaments are expected to collapse and fragment. The subcritical filaments bounce and disperse.}

\begin{figure*}
  \centering
  \includegraphics[width=0.45\textwidth]{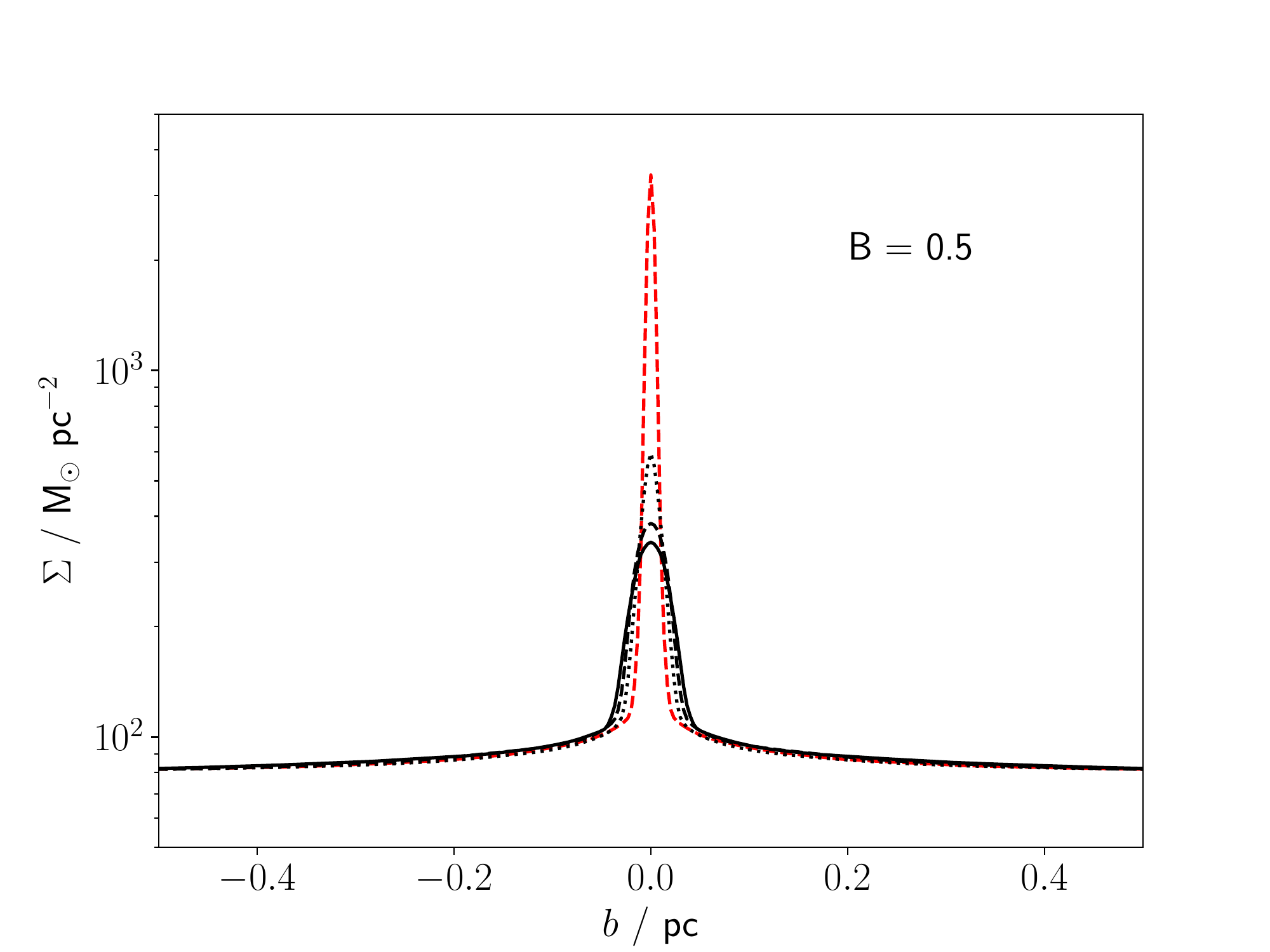}\quad
  \includegraphics[width=0.45\textwidth]{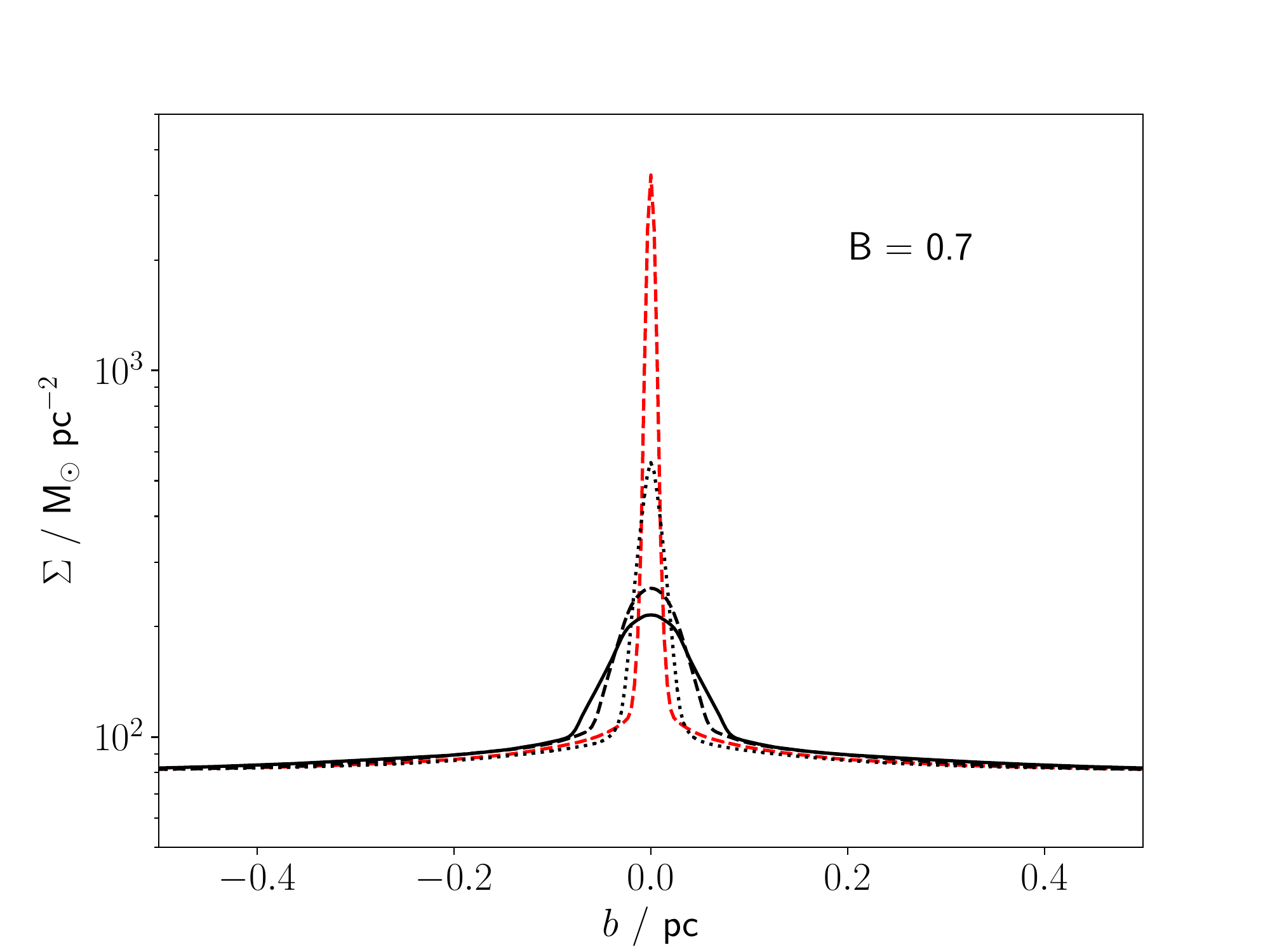}\quad
  \caption{{Surface-density profiles for supercritical models with ${\cal M}=3$, simulated using ideal MHD, and viewed after $1.09 \myr$, at an angle of $\phi=0^\circ$ (solid black lines), $45^\circ$ (dashed black lines) and $90^\circ$ (dotted black lines). $\,\phi$ is the angle between the initial magnetic field and the line-of-sight. $\,b$ (the abscissa) is the impact parameter of the line-of-sight with respect to the filament spine. {\it Left panel:} ${\cal B}\subperp=0.5$. {\it Right panel:} ${\cal B}\subperp=0.7$.} Red dashed lines show the equivalent profile for a model with ${\cal B}\subperp=0$ (no magnetic field). A background surface-density of $80 \msun \pc^{-2}$ ($N\submolH\sim4.2\times 10^{21}\,\rm{cm}^{-2}$) has been added to the profiles.}
  \label{fig:sdens}
\end{figure*}

\begin{figure*}
  \centering
  \includegraphics[width=0.45\textwidth]{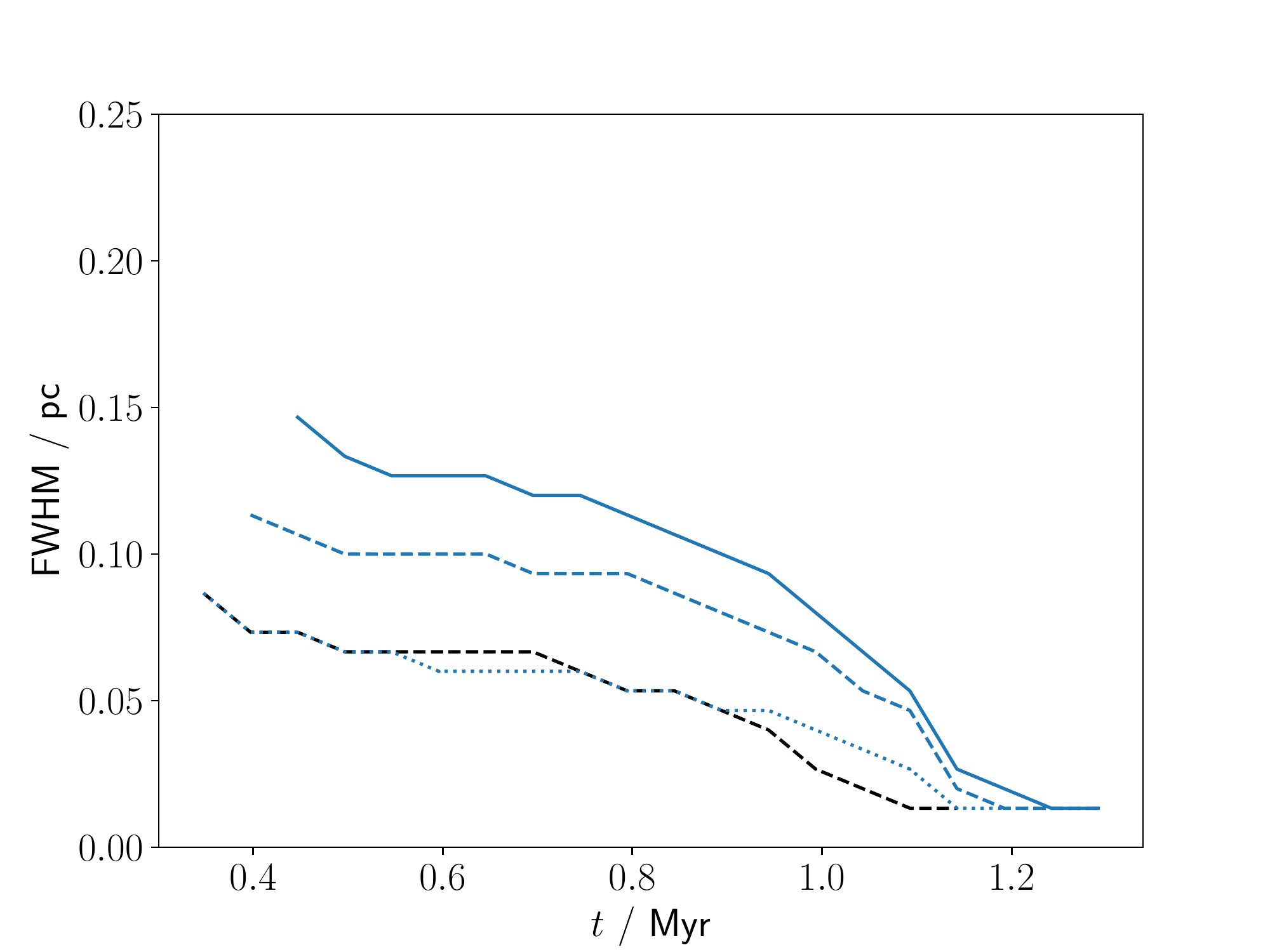}\quad
  \includegraphics[width=0.45\textwidth]{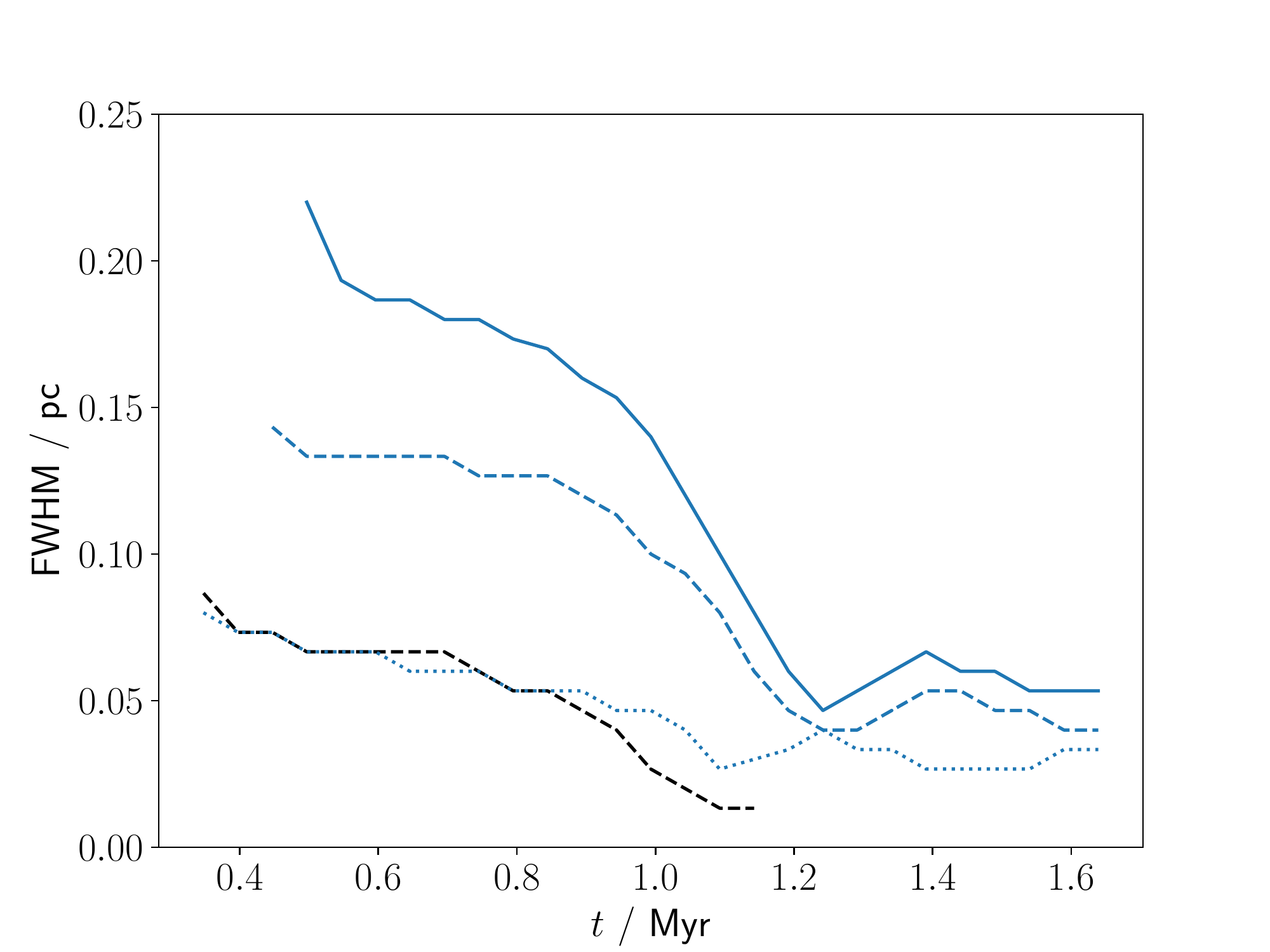}\quad
  \caption{{The evolution of the filament $\fwhm$ for supercritical models with ${\cal M}=3$, simulated using ideal MHD, and viewed at an angle of $\phi=0^\circ$ (solid blue lines), $45^\circ$ (dashed blue lines) and $90^\circ$ (dotted blue lines); $\phi$ is the angle between the initial magnetic field and the line-of-sight. {\it Left panel:} ${\cal B}\subperp=0.5$. {\it Right panel:} ${\cal B}\subperp=0.7$.} The dashed black line shows the evolution for ${\cal B}\subperp=0$.}
  \label{fig:width}
\end{figure*}

\section{Model}

Our idealised converging-flow model of filament formation is described in detail in \citet{priestley2022}. Here we briefly summarise the main elements, {and how the model is extended to include a magnetic field.} The gas that is destined to form the filament is isothermal with sound speed $a\subO$, and initially it is contained in a cylinder of radius $W\subO$ and length $L\subO$, with uniform density, $\rho\subO$, and uniform radial inward velocity, $v\subO$. This cylinder is surrounded by stationary gas with lower density, $\rho\subO/10$, and higher isothermal sound speed, $10^{1/2}a\subO$, which fills the rest of the computational domain. The symmetry axis of the filament is the $z$-axis. For the non-magnetic model there are only two fundamental parameters, both dimensionless, (i) the ratio of the line-density to the critical value for gravitational stablility \citep{StodolkiewiczJ1963,ostriker1964}, ${\cal G}=\pi GW\subO^2\rho\subO/2a\subO^2$, and (ii) the Mach number of the inflow, ${\cal M}=v\subO/a\subO$. To make the numerical results specific, we set $L\subO =5\pc$, $W\subO=1\pc$ and $a\subO=0.187\kms$ (corresponding to molecular gas with solar composition at $T=10\kel$).

For the {models} presented here we fix ${\cal G}=1.2$ {(and hence $\rho\subO=6.2\msun\pc^{-3}$, $n\submolHO=88\,\rm{cm^{-3}}$), but we vary ${\cal M}$. Additionally,} in order to specify the initial perpendicular magnetic field, $B\subOperp$, we introduce a third fundamental dimensionless parameter,
\begin{eqnarray}
{\cal B}\subperp&=&\frac{2\,[5/G]^{1/2}B\subOperp}{3\pi^2W\subO\rho\subO},
\end{eqnarray}
which is the ratio of the magnetic flux through the filament to the critical value. {Thus supercritical filaments have ${\cal B}\subperp<1$, and subcritical filaments have ${\cal B}\subperp>1$.} It follows that
\begin{eqnarray}\label{EQN:BO.02}
B\subOperp&=&\frac{3\pi a\subO^2 {\cal G} {\cal B}\subperp}{(5G)^{1/2} W\subO}\;\;=\;\;2.22\ug\,\left[\frac{\cal G}{1.2}\right]\,{\cal B}\subperp.
\end{eqnarray}
The initial magnetic field is aligned with the $y$-axis.

{For ${\cal B}\subperp \sim 1$, the initial field strengths are much lower than the estimated fields in observed molecular filaments ($\sim 10-100 \ug$; \citealt{pattle2021,lyo2021}). However, these weak initial fields simply reflect our arbitrary choice of $W\subO$; they are rapidly amplified to much higher values by the subsequent converging flows. For example, in the simulation with ${\cal M} = 3$ and ${\cal B}\subperp = 0.5$, the initial field strength is $\sim 1 \ug$, but by the end of the simulation it exceeds $1000 \ug$ in the centre of the filament. The low initial field strengths of our models are thus entirely consistent with the observed values, and simply reflect that fact that we have started the simulations from rather diffuse initial conditions.}

The simulations are performed with the {\sc phantom} Smoothed Particle (Magneto-)Hydrodynamics (SPH) code \citep{price2018}, using $5 \times 10^6$ particles and the default smoothing-length coefficient $h\subfact=1.2$. Hence an SPH particle has mass $m\subSPH=2 \times 10^{-5} \msun$, and the notional mass resolution is $\sim58\,m\subSPH\sim1.2\,{\rm M}_{\rm Jup}$. We terminate the simulations once the density reaches $\rho\subMAX\simeq1.5 \times 10^7\msun\pc^{-3}$. Beyond this point the filament starts to fragment into prestellar cores \citep{clarke2016}, and so there is no longer a representative filament profile to speak of. We construct average {surface-density} profiles using the central section of the filament, $|z| \le 1 \pc$, to avoid {the parts of the filament that are affected by} end-dominated longitudinal collapse \citep{PonAetal2012b,clarke2015}.

\section{Results}

\subsection{Supercritical filaments (${\cal B}\subperp < 1$)}

{We simulate supercritical filaments with ideal MHD, rather than non-ideal MHD, because this requires much less computing resource, and non-ideal effects are small --- as shown in Appendix \ref{sec:nimhd}, where we evaluate the ambipolar diffusion timescale analytically, and Appendix \ref{APP:NonIdeal} where we evolve one of the models using non-ideal MHD, and show that the resulting distribution of $\fwhm$s is not significantly changed from the results obtained with ideal MHD. {We note parenthetically that \citet{HennebellePAndreP2013} also conclude that non-ideal effects can safely be ignored in modelling the assembly of a filament.}}

Figure \ref{fig:sdens} shows surface-density profiles for ${\cal M} = 3$ (i.e. an intermediate {inflow} velocity, $v\subO\sim0.56\,\rm{km\,s^{-1}}$) and ${\cal B}\subperp=0.5$ (left  panel) or ${\cal B}\subperp=0.7$ (right panel). {The profiles are computed on the assumption that the filament spine (the $z$-axis) lies on the plane of the sky, and therefore they depend on the angle of the line-of-sight with respect to the magnetic field direction (the $y$-axis).} Compared with the non-magnetic case, the peak densities are lower, and the filament profiles are broader, due to the added magnetic support. {However, since the mass-to-flux ratio exceeds the critical value (i.e. the filaments are still supercritical), this additional support only delays collapse, rather than preventing it.} The visible effects of the field are strongest when the filament is viewed along the magnetic-field direction (i.e. along the $y$-axis); and weakest when the filament is viewed perpendicular to the magnetic-field direction (i.e. along the $x$-axis); {in the latter case,} the widths are very similar to the non-magnetic case.

As in \citet{priestley2022}, we only consider the filament $\fwhm$s when {two conditions are met. First, the peak of the surface-density on the spine must be at least three times greater than that of the initial profile, i.e. $\Sigma\subSPINE>\Sigma\subMIN\simeq37.2 \msun \pc^{-2}$. Below this threshold, the filament would be poorly defined \citep[cf.][]{arzoumanian2019}, and would probably not be detected by an automated filament finding algorithm. Second, the volume-density on the spine of the filament, $\rho\subSPINE$, must be below $\rho\subMAX\simeq1.5 \times 10^7\msun\pc^{-3}$. As noted above, the simulations are terminated at this point, because thereafter the filament starts to fragment, and therefore the surface-density profile is not well defined. In the sequel we refer to filaments satisfying these two conditions ($\Sigma\subSPINE>\Sigma\subMIN$, $\,\rho\subSPINE<\rho\subMAX$) as {\it observable filaments}.}

Figure \ref{fig:width} shows the time evolution of the $\fwhm$ for observable filaments from the models with ${\cal M}=3$,  and ${\cal B}\subperp=0.5$ {(left panel)} or $0.7$ {(right panel)}. For comparison we also show the model with ${\cal M}=3$ and ${\cal B}\subperp=0$ (no magnetic field; {black dashed line)}. For a viewing angle of $90^\circ$ to the $y$-axis, {the $\fwhm$s of the magnetic cases evolve} very similarly to the non-magnetic case, as there is no magnetic {resistance to} matter flowing along the field lines. However, {the additional magnetic resistance to matter flowing perpendicular to the field lines} means that magnetised models then take longer to reach $\rho\subMAX$. {Consequently, for viewing angles close to $90^\circ$,} the magnetic models stall {briefly} at small values of $\fwhm$. For small viewing angles relative to the $y$-axis, the $\fwhm$ is larger at any given time, and goes from being significantly larger than the non-magnetic value at early times ($\lesssim 0.4\myr$) to being comparable at the end of the simulation ($\gtrsim 1.1\myr$).

\begin{figure*}
  \centering
  \includegraphics[width=0.3\textwidth]{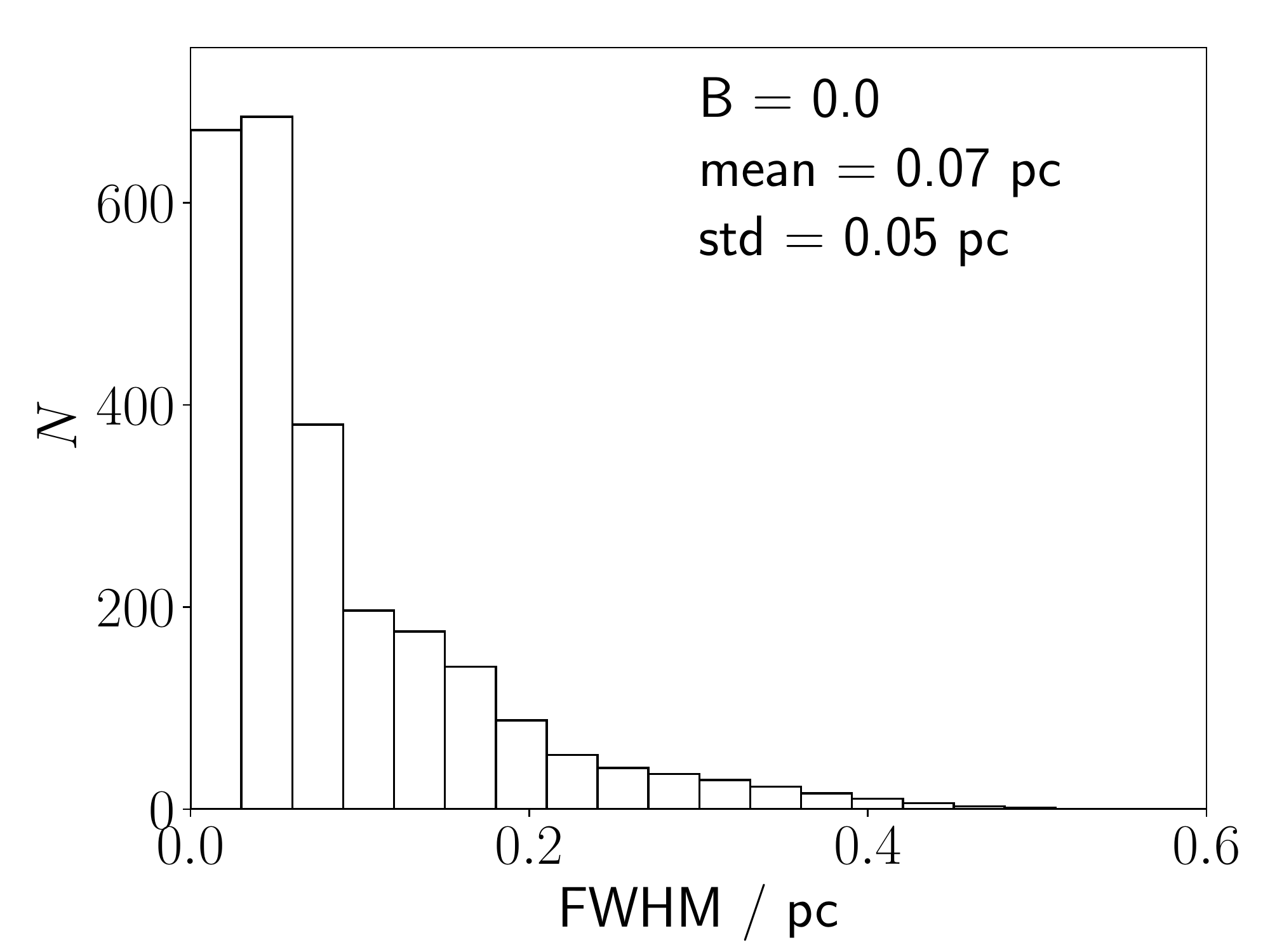}\quad
  \includegraphics[width=0.3\textwidth]{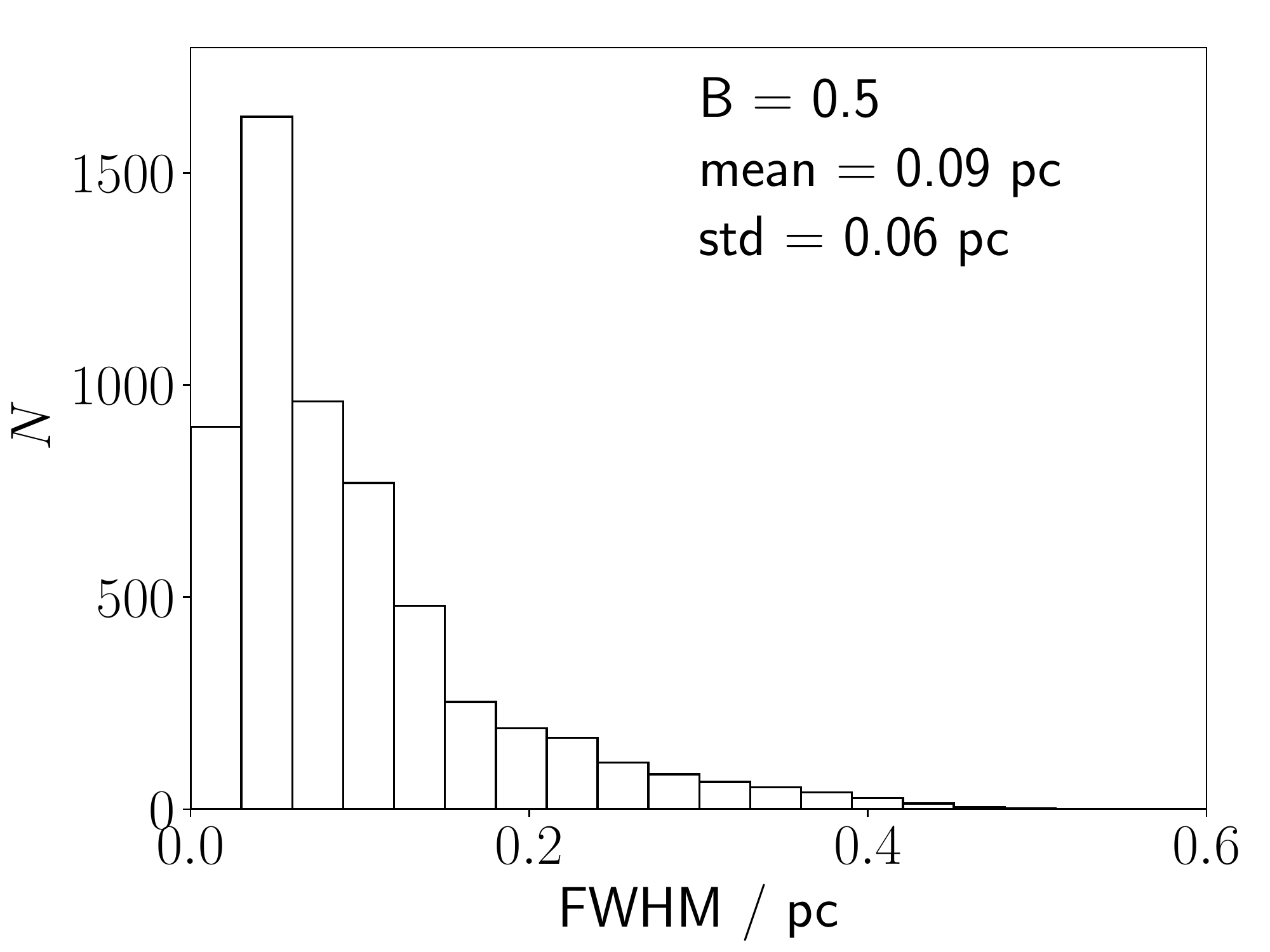}\quad
  \includegraphics[width=0.3\textwidth]{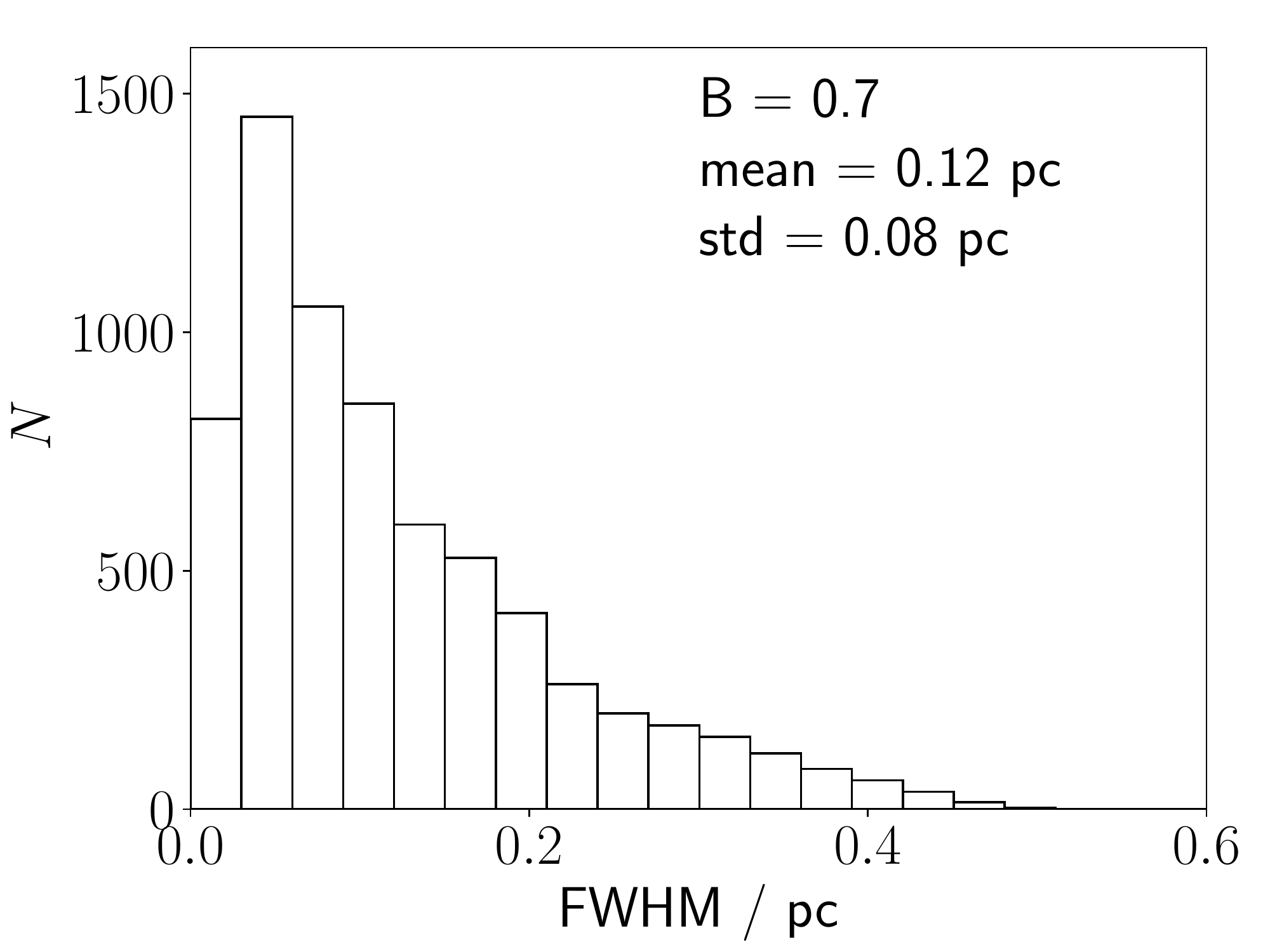}\quad
  \caption{The distribution of filament $\fwhm$s for supercritical models with a flat distribution of inflow Mach Numbers, simulated using ideal MHD, and viewed from a flat distribution of viewing angles, $\phi$. {\it Left panel:} ${\cal B}\subperp=0$. {\it Centre panel:} ${\cal B}\subperp=0.5$. {\it Right panel:} ${\cal B}\subperp=0.7$.}
  \label{fig:hist}
\end{figure*}

\begin{figure*}
  \centering
  \includegraphics[width=\columnwidth]{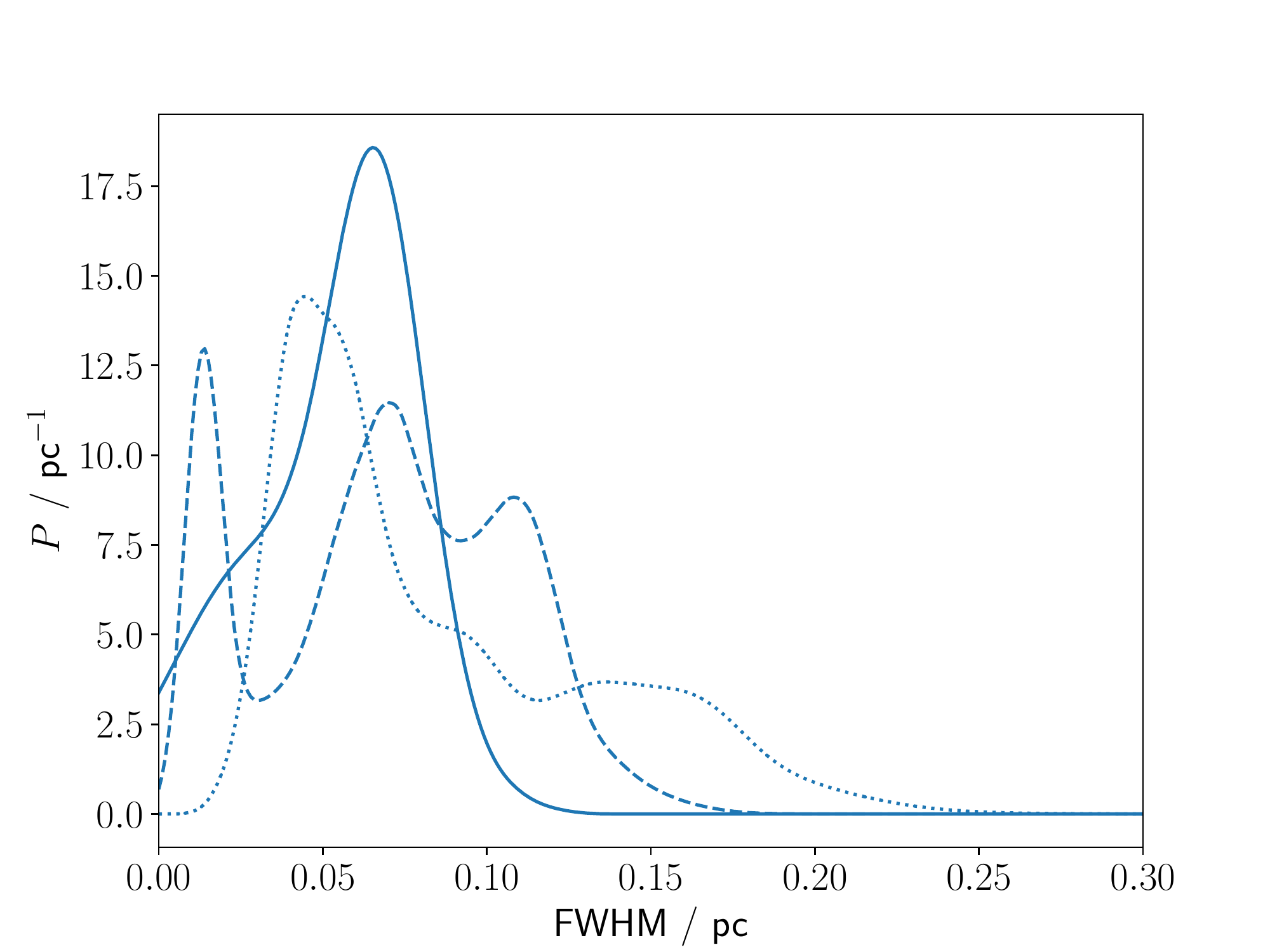}\quad
  \includegraphics[width=\columnwidth]{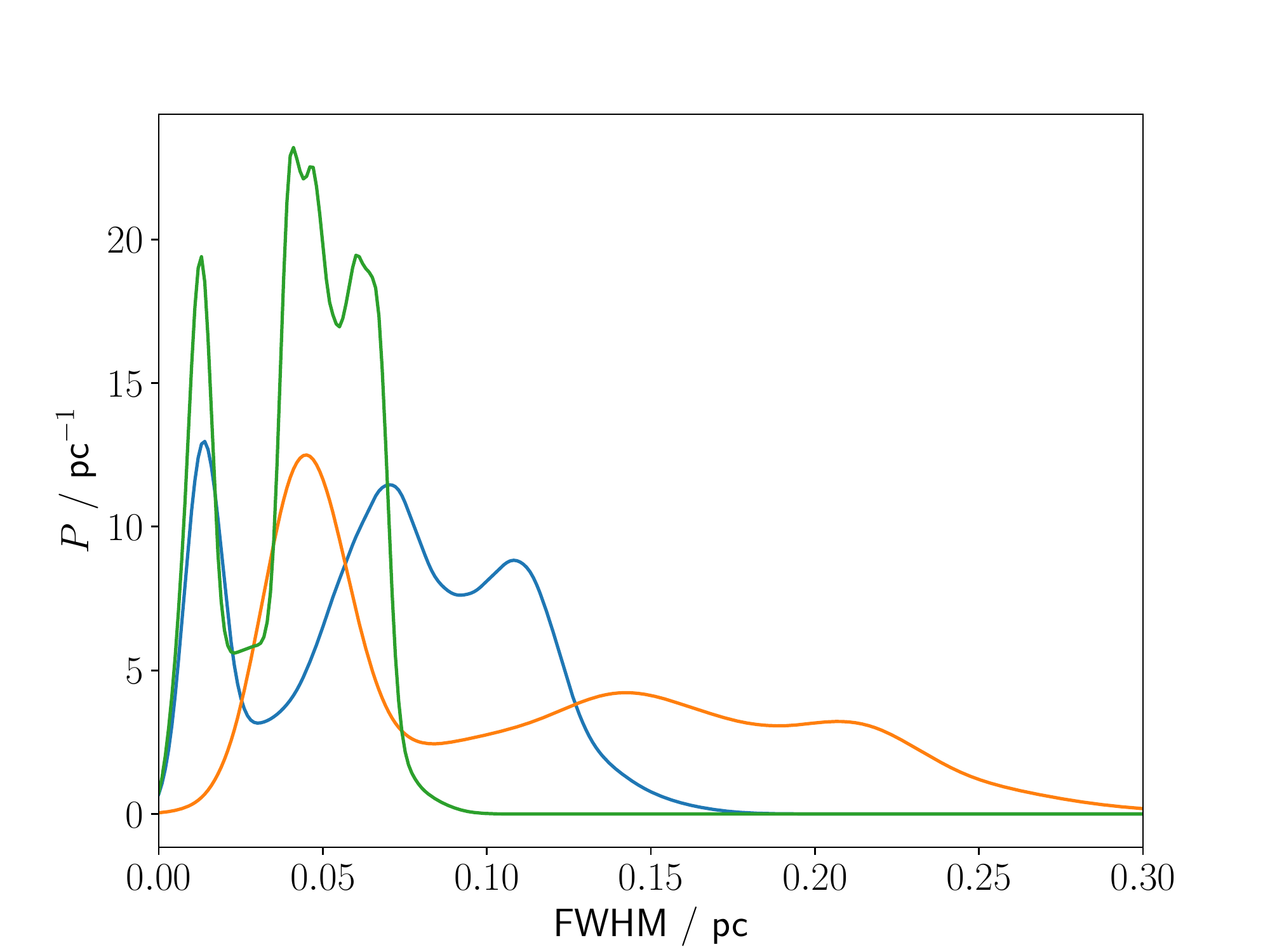}
  \caption{The distributions of filament $\fwhm$s for supercritical models with specific combinations of inflow Mach Number, ${\cal M}$, and magnetic-field parameter, ${\cal B}\subperp$, simulated using ideal-MHD and observed from a flat distribution of viewing angles, $\phi$. {\it Left panel:} Models with ${\cal M}=3$ and ${\cal B}\subperp=0$ (solid line), $0.5$ (dashed line), and $0.7$ (dotted line). {\it Right panel:} Models with ${\cal B}\subperp = 0.5$ and ${\cal M}=2$ (orange), $3$ (blue), and $4$ (green).}
  \label{fig:pdf}
\end{figure*}

{In order to generate distributions of observable filament $\fwhm$s, we make three assumptions. First, we assume that an observable filament is equally likely to be observed at any time, i.e. between when its peak surface-density first exceeds $\Sigma\subMIN \simeq 37.2 \msun \pc^{-2}$ and when its maximum volume-density first exceeds $\rho\subMAX \simeq 1.5 \times 10^7\msun\pc^{-3}$. Second, we assume that the spine of the observable filament is on the plane of the sky. Therefore the observable filament's aspect only depends on the angle, $\phi$, between the viewing direction and the initial magnetic field direction (the $y$ axis), with $\phi$ distributed uniformly between $\phi =0^{\circ}$ and $\phi =90^{\circ}$. We term this a flat distribution of viewing angles.\footnote{In reality filament axes will be tilted relative to the viewing direction. This will increase the surface-density, making filaments `more observable', but given the sharp edges of the modelled filaments \citep[see][]{priestley2022}, this will not significantly affect the distribution of observed $\fwhm$s. It will also shorten apparent (i.e. projected) filament lengths, so that some filaments aligned close to the line-of-sight will be too short to be identified as filaments \citep[e.g.][]{arzoumanian2019}, but again this is a small effect.} Third, when we generate distributions of observable filament $\fwhm$s for an ensemble of inflow Mach Numbers, ${\cal M}$, we assume that ${\cal M}$ is distributed uniformly over the range $0.5\lesssim{\cal M}\lesssim5.5$; we term this a flat distribution of inflow Mach Numbers.}

{Figure \ref{fig:hist} shows histograms of the observable filament $\fwhm$s obtained for models with ${\cal B}\subperp=0$ (left panel), $0.5$ (middle panel) and $0.7$ (right panel), and a flat distribution of inflow Mach Numbers, ${\cal M}$. As expected, the mean and the standard deviation both increase with increasing ${\cal B}\subperp$. However, the shape of the distribution does not change. There is a peak at small $\fwhm\sim0.04(\pm0.01)\,\rm{pc}$ and a power-law tail to higher values, similar to what is observed \citep[][]{panopoulou2017}. This is despite the introduction of a perpendicular magnetic field causing significant changes to the $\fwhm$ distribution for any particular value of ${\cal M}$, shown in the left panel of Figure \ref{fig:pdf}. Depending on the value of ${\cal B}\subperp$, the $\fwhm$ distribution for ${\cal M} = 3$ may be multi-peaked, and extend to smaller or larger $\fwhm$s than the non-magnetised case. However, the range of the $\fwhm$ distribution is much more strongly affected by ${\cal M}$, as shown in the right panel of Figure \ref{fig:pdf}. Thus, while the $\fwhm$ distribution {\it for a single value of ${\cal M}$} may appear to differ from those observed, when considering the average distribution {\it over a range of ${\cal M}$}, the observed single-peaked behaviour is recovered. That is, as long as the filament mass-to-flux ratio is {\it supercritical} (${\cal B}\subperp < 1$).}

\begin{figure*}
  \centering
  \includegraphics[width=0.45\textwidth]{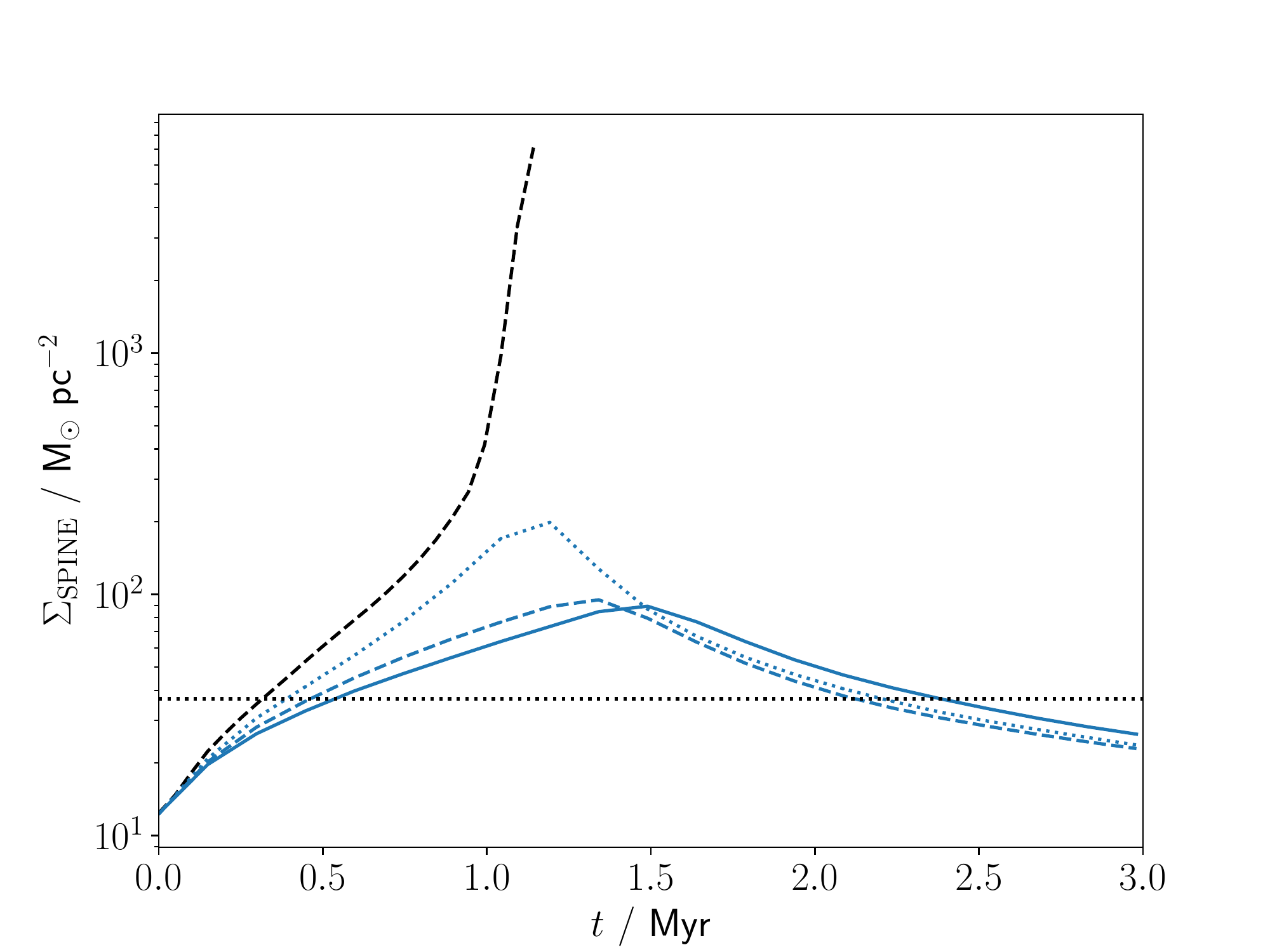}\quad
  \includegraphics[width=0.45\textwidth]{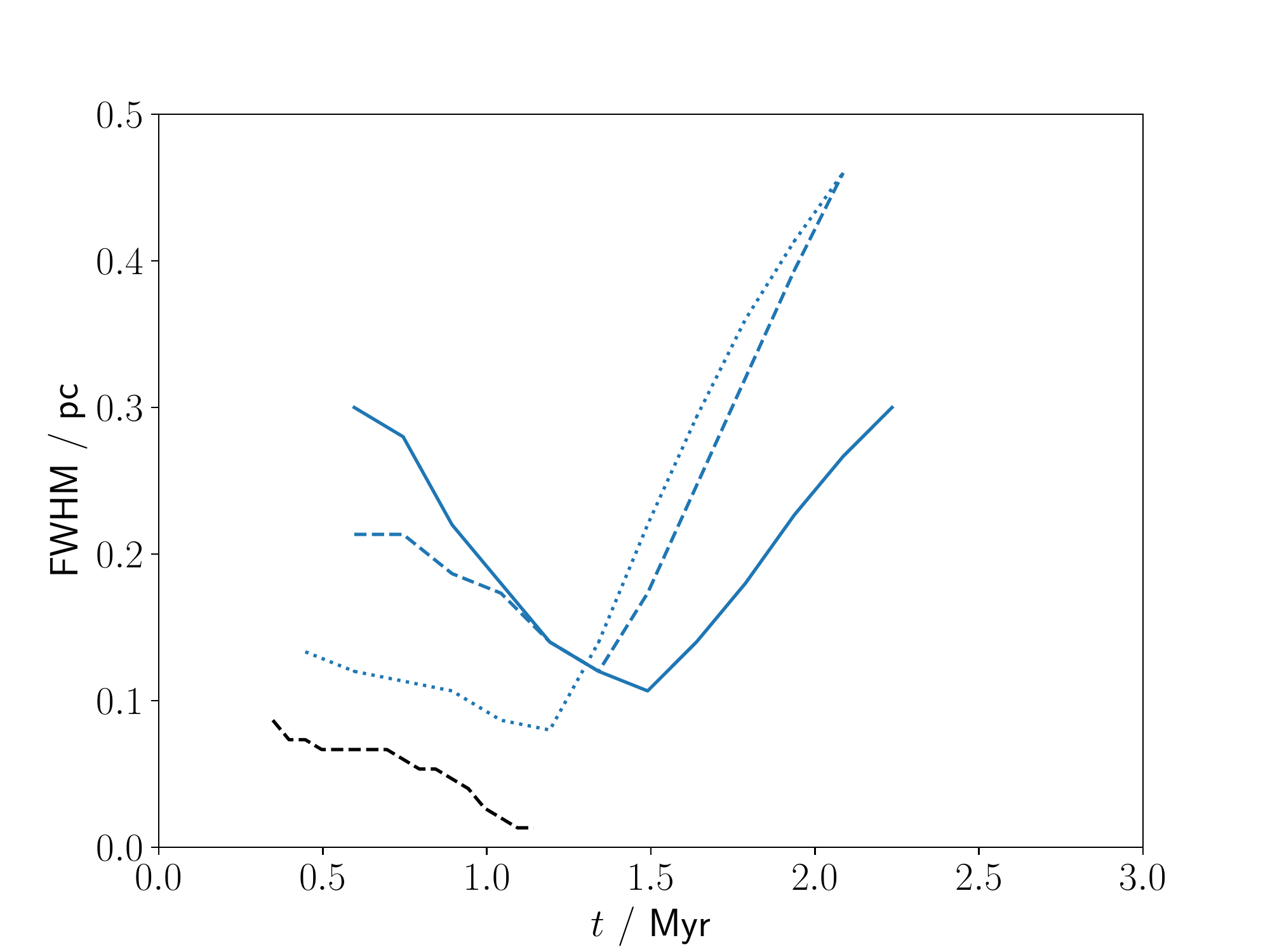}\quad
  \caption{{Subcritical models with ${\cal M}=3$ and ${\cal B}\subperp=1.5$, simulated using non-ideal MHD, and viewed at an angle of $\phi=0^\circ$ to the field direction (solid blue lines), $45^\circ$ (dashed blue lines) and $90^\circ$ (dotted blue lines). {\it Left panel:} evolution of the spinal surface-density, $\Sigma\subSPINE$. {\it Right panel:} evolution of the filament $\fwhm$.} The dashed black lines show the evolution for ${\cal B}\subperp=0$. {On the left panel,} the dotted, horizontal black line shows the surface-density threshold of $\Sigma\subMIN=37.2 \msun \pc^{-2}$.}
  \label{fig:widthsub}
\end{figure*}

\begin{figure}
  \centering
  \includegraphics[width=0.45\textwidth]{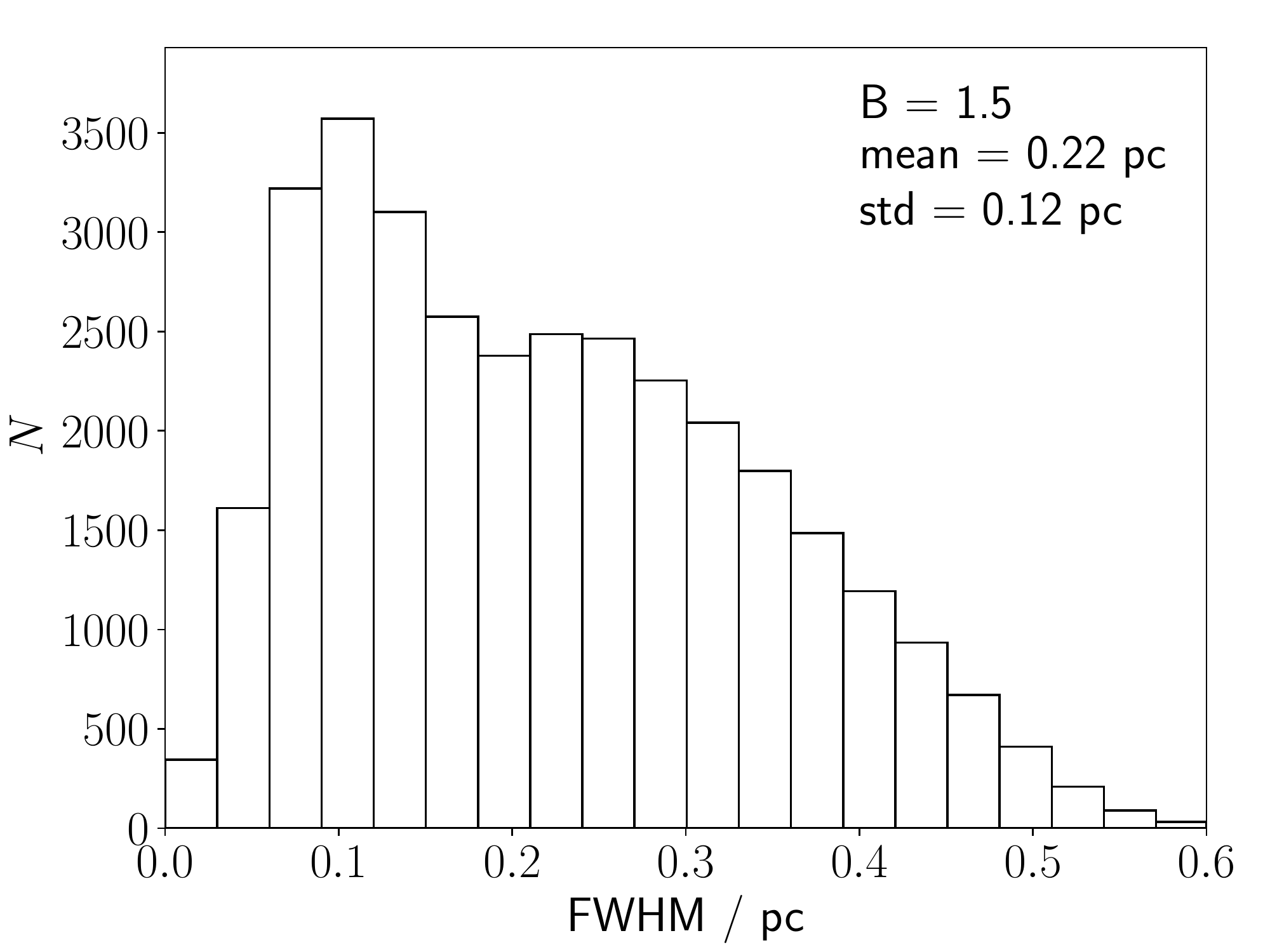}\quad
  \caption{The distribution of filament $\fwhm$s for subcritical filaments with ${\cal B}\subperp =1.5$ and a flat distribution of inflow Mach Numbers, simulated using non-ideal MHD and viewed from a flat distribution of viewing angles.}
  \label{fig:histsub}
\end{figure}

\subsection{Subcritical filaments (${\cal B}\subperp > 1$)}\label{SEC:SubFils}

{A filament with subcritical mass-to-flux ratio (${\cal B}\subperp > 1$) is prevented from collapsing under self-gravity, and so, if evolved with ideal MHD, never reaches the maximum density ($\rho\subMAX\simeq1.5 \times 10^7\msun\pc^{-3}$) that we use as a stopping point.\footnote{Due to the finite filament length, our models will eventually reach this density via longitudinal contraction, but this does not affect the arguments presented here.} Instead, such a filament alternates between periods of expansion and contraction as it attempts to reach a stable configuration. Non-ideal MHD processes will eventually allow the filament's self-gravity to overcome magnetic support, but on a much longer timescale than typical filament lifetimes (see Appendix \ref{sec:nimhd}). We therefore terminate these models after $3 \myr$, which is the maximum filament-compression timescale, as estimated by \citet{auddy2016}. Molecular clouds are thought to have lifetimes at most a few times longer than this \citep[e.g.][]{chevance2022}, so subcritical filaments cannot evolve indefinitely without being disrupted by feedback from ongoing nearby star formation.}

{We account for non-ideal MHD processes in subcritical models using the NICIL library \citep{wurster2016}, with the ionisation fraction, $x_{\rm i}$, given by
\begin{eqnarray}\label{EQN:xi}
x_{\rm i}&\equiv&\frac{n_{\rm i}}{\nh}\;\;=\;\;6\times 10^{-6} \left(\frac{\nh}{\pcc}\right)^{-0.6}.
\end{eqnarray}
Here $n_{\rm i}$ is the volume-density of ions and $\nh$ is the volume-density of hydrogen in all chemical forms. Equation~\ref{EQN:xi} accurately reproduces the results of time-dependent chemical models \citep{tassis2012b,priestley2019}. The most important non-ideal process under the physical conditions with which we are concerned here is ambipolar diffusion \citep{wurster2021}. However, even ambipolar diffusion has a characteristic timescale much longer than the lifetimes of filaments and molecular clouds, as shown in Appendix \ref{sec:nimhd}.}

{Figure \ref{fig:widthsub} shows the evolution of the maximum surface-density and the $\fwhm$ for ${\cal M} = 3$ and ${\cal B}\subperp = 1.5$. As with supercritical models, when viewed at $90^\circ$ to the magnetic field direction, the filament evolution is initially close to the ${\cal B}\subperp = 0$ case. However, the collapse is eventually halted by pressure forces, and the filament then begins to expand, resulting in a drop in the maximum surface-density and an increase in $\fwhm$. Despite the inclusion of non-ideal processes, the filament does not begin to contract again during the $3 \myr$ over which we follow its evolution. As shown in Appendix \ref{sec:nimhd}, significant ambipolar diffusion occurs on a much longer timescale. While the filament is contracting, it appears denser and narrower when viewed with the magnetic field close to the plane of the sky (i.e. from directions, $\phi\sim90^\circ$), since the gas can flow freely {\it inwards} along the field lines. However, once the filament `bounces' and begins expanding, the gas flows freely {\it outwards} along the field lines, and so the filament then appears broader when viewed with the magnetic field close to the plane of the sky (see right panel of Figure \ref{fig:widthsub}).}

{At times and viewing angles where a subcritical filament is detectable, its $\fwhm$ is typically much larger than the corresponding supercritical models. Figure \ref{fig:histsub} shows the resulting $\fwhm$ distribution for ${\cal B}\subperp=1.5$ and a flat distribution of inflow Mach Numbers. It is still approximately single-peaked, but has a very different shape to the distributions for supercritical models (shown in Figure \ref{fig:hist}), being much broader, and having a substantial fraction of filaments with $\fwhm > 0.2 \pc$. The mean and median filament $\fwhm$s are both $\sim 0.2 \pc$, comparable to the values from the analytical model of \citet{auddy2016}, although the range of $\fwhm$s in our models is significantly greater.}

\section{Discussion}

{When $\fwhm$s are evaluated at individual points along the spine of an observed filament -- rather than being derived from a single profile, averaged along the length of the filament -- the resulting distribution of $\fwhm$s has three key properties \citep{panopoulou2017,arzoumanian2019}. (i) The distribution shows a distinct peak at $\fwhm \lesssim 0.1 \pc$. (ii) The distribution falls off sharply (approximately as a power law) to larger $\fwhm$s. (iii) The $\fwhm$s are essentially uncorrelated with the surface-density on the spine \citep{arzoumanian2019}. We should therefore compare our models with these key observational properties.}

\subsection{The peak of the distribution of $\fwhm$s}\label{SEC:PDFPeak}

{For supercritial values of ${\cal B}\subperp$, the $\fwhm$ distributions presented in Figure \ref{fig:hist} all have peaks in the range $\sim 0.03\pc$ to $\sim 0.05\pc$, somewhat below the peak at $\sim 0.1 \pc$ derived by \citet{arzoumanian2011,arzoumanian2019}. However, the peak at $\sim 0.1 \pc$ has recently been questioned by \citet{panopoulou2022}, based on updated distances to molecular clouds obtained} using {\it Gaia} data. \citet{panopoulou2022} find that the {estimated} average filament $\fwhm$ in a molecular cloud increases with distance. {This strongly suggests} that unappreciated resolution effects have been overlooked. {Moreover,} the four molecular clouds in their sample which fall within $200 \pc$, and are therefore presumably least affected by resolution issues, have a consistent averaged filament $\fwhm$ of $\sim0.05 \pc$, in {good} agreement with {the results presented} in Figure \ref{fig:hist}.

Studies based on higher-resolution observations than the {\it Herschel} data used by \citet{arzoumanian2011,arzoumanian2019} have found filament $\fwhm$s below $0.1 \pc$ (e.g. \citealt{hacar2018,li2022}). These {estimates are mostly} based on line emission from molecules such as HCO$^+$ and N$_2$H$^+$, which will almost inevitably return $\fwhm$s {lower than the true value,} because they only trace the densest filament material \citep{priestley2020}. {However, we note that \citet{li2022} have detected one filament for which the dust continuum emission gives $\fwhm \sim 0.03 \pc$. Given the inherent uncertainties, the peak of the distribution of filament $\fwhm$s predicted by our supercritical filament formation model is encouragingly close to the observational estimates.}

{The distribution of $\fwhm$ for our subcritical models (see Figure \ref{fig:histsub}) peaks around the $\sim 0.1 \pc$ value found by \citet{arzoumanian2011,arzoumanian2019}, but the peak is substantially broader than the observed distributions, and extends to much larger values. Whereas an intrinsically narrow filament might appear broader if viewed with limited resolution, no credible effect exists which would make an intrinsically broad filament appear narrower. Thus, if the true distribution of filament $\fwhm$s really did extend as strongly beyond $\sim 0.2 \pc$ as in Figure \ref{fig:histsub}, this would be immediately apparent in the data. In fact, as we discuss below, the fraction of observed filaments with $\fwhm$s beyond $\sim 0.2 \pc$ is small enough to be negligible \citep{panopoulou2017,arzoumanian2019}. The analytical model of \citet{auddy2016} predicts a mean $\fwhm$ of $\sim 0.3 \pc$ for subcritical filaments formed by compression, comparable with the average values of our model distribution. This strongly suggests that broad $\fwhm$s are a universal property of observable subcritical filaments. It would appear that the observations cannot be reconciled with subcritical filament formation.}

\begin{figure*}
  \centering
  \includegraphics[width=0.3\textwidth]{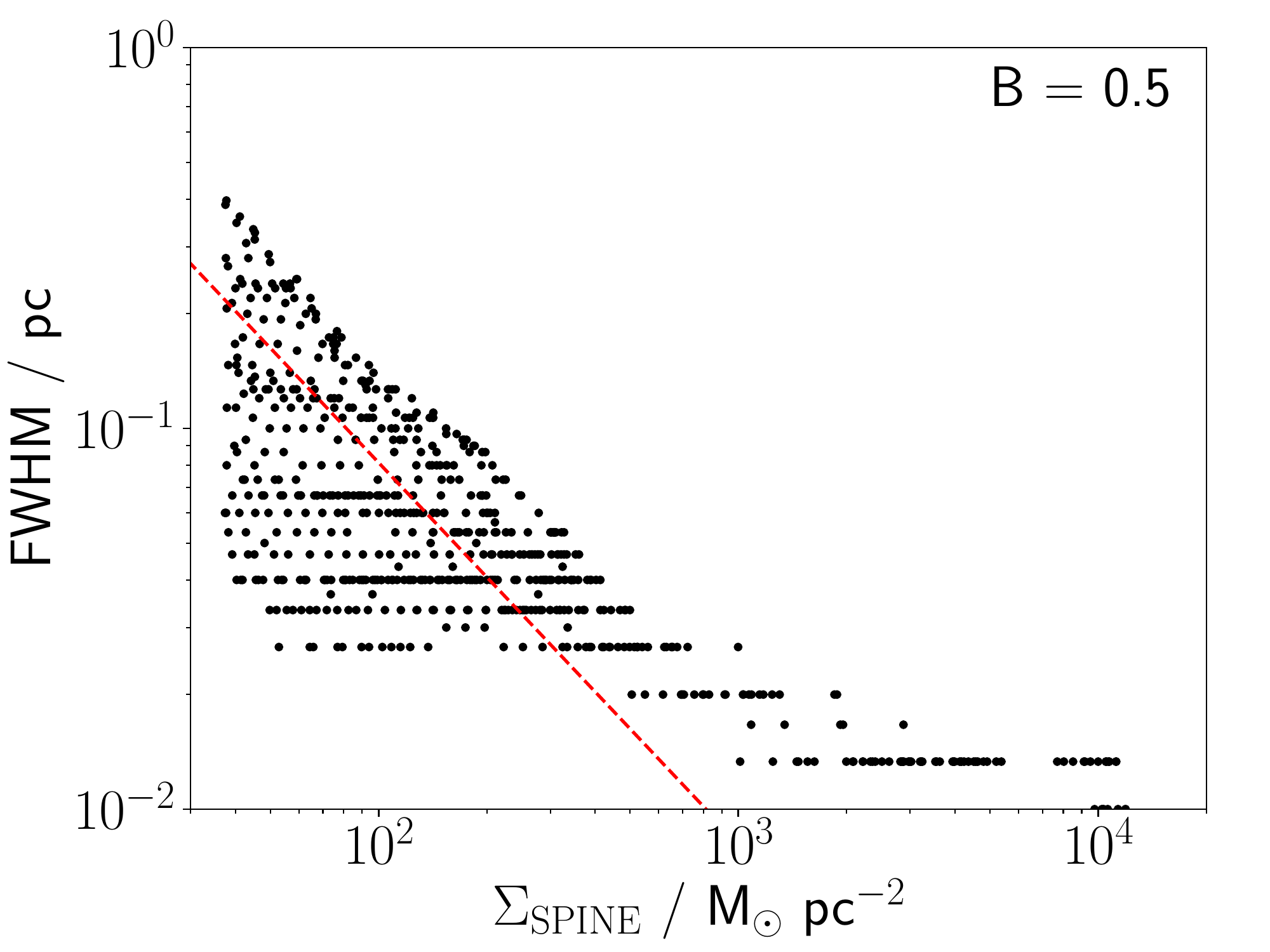}\quad
  \includegraphics[width=0.3\textwidth]{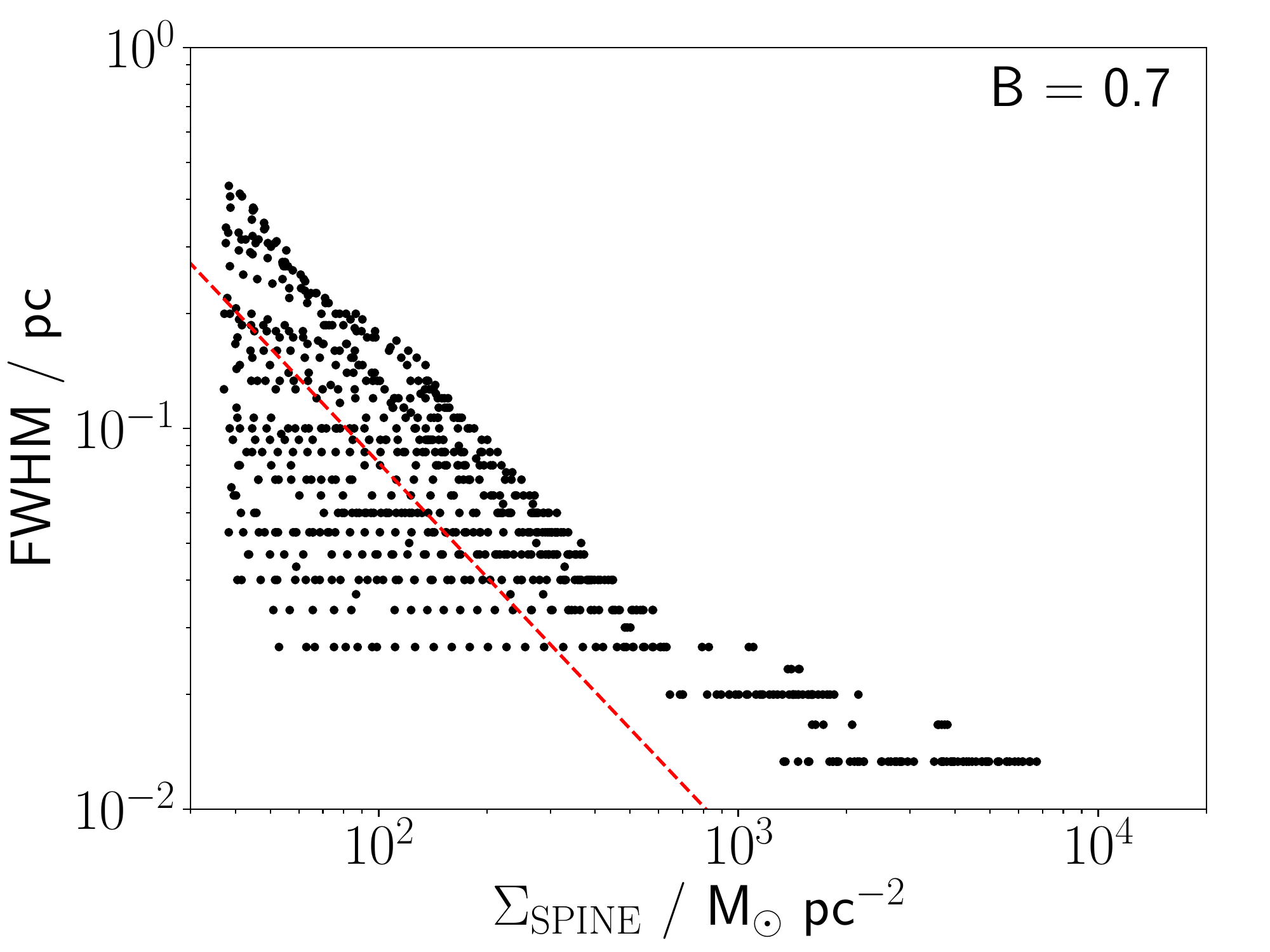}\quad
  \includegraphics[width=0.3\textwidth]{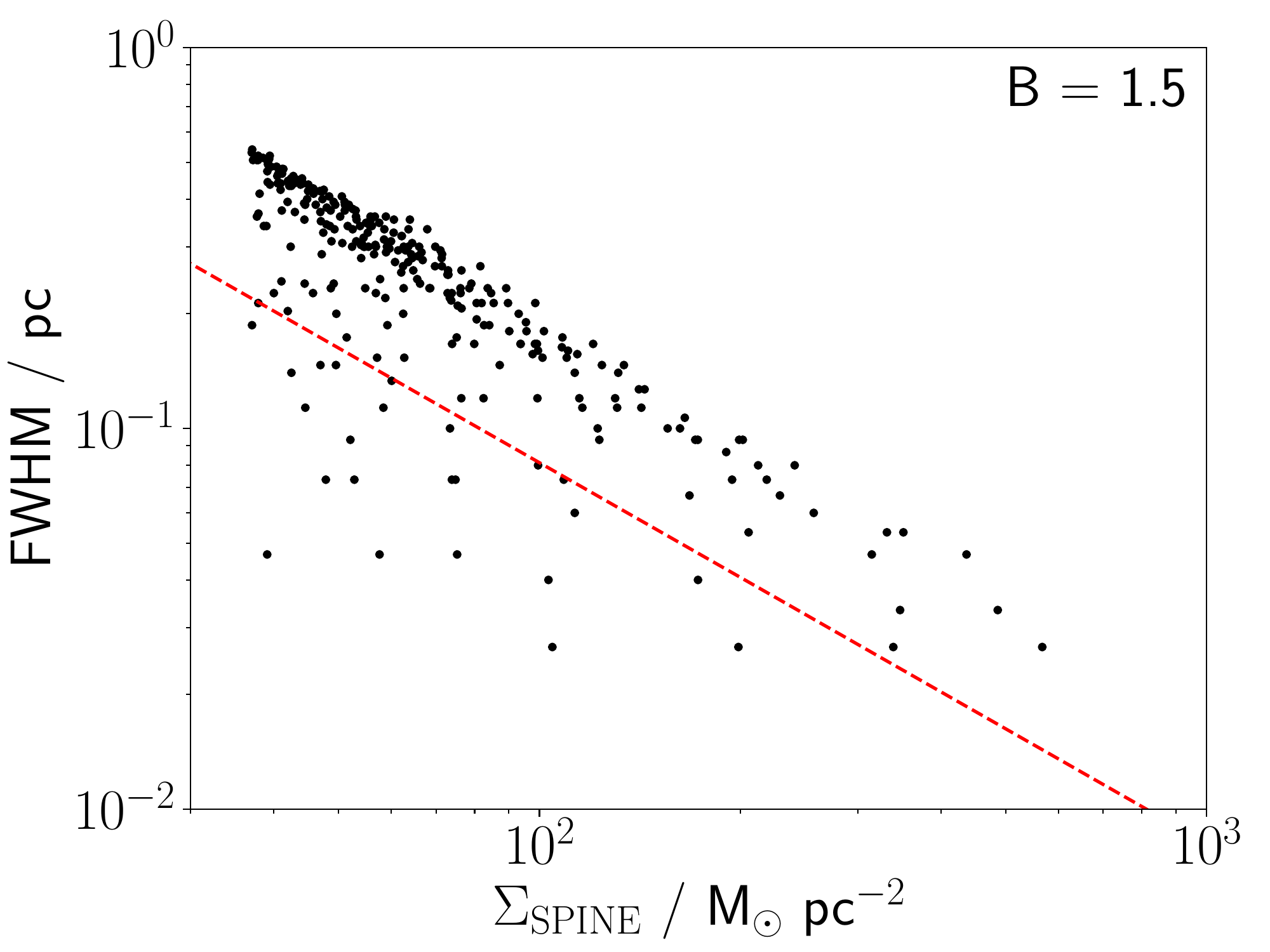}\quad
  \caption{Full-width at half-maximum, $\fwhm$, versus spinal surface-density, $\Sigma\subSPINE$, for models with ${\cal B}\subperp = 0.5$ (left panel; simulated using ideal MHD), $0.7$ (central panel; simulated using ideal MHD), and $1.5$ (right panel; simulated using non-ideal MHD). The dashed red lines indicate the Jeans length on the spine, $\lambda\subJEANS$ (see Equation~\ref{EQN:lambdaJEANS}).}
  \label{fig:fwsd}
\end{figure*}

\subsection{The proportion of filaments with high $\fwhm$}\label{SEC:FatFils}

The sharp power-law decline in the number of filaments with $\fwhm$s greater than the peak value (when no averaging along the length of the filament is performed; \citealt{panopoulou2017,arzoumanian2019}) places a second constraint on models of filament formation. Our {supercritical models naturally produce a power-law decline, $\sim\fwhm^{-1.5}$,} but this is much shallower than the $\fwhm^{-5}$ decline derived by \citet{panopoulou2017} from the observations, and somewhat shallower than the $\fwhm^{-2.6}$ decline derived by \citet{arzoumanian2019} from the observations. {Even with ${\cal B}\subperp = 0$, we find that the percentage of model filaments with $\fwhm >0.2 \pc\,$ is $\,f\subbroad\sim 8\%$. For finite but {\it supercritical} magnetic fields, we find $f\subbroad\sim 10\%$ for ${\cal B}\subperp=0.5$, and $f\subbroad\sim16\%$ for ${\cal B}\subperp=0.7$. These percentages must be contrasted with the much smaller percentage in the data analysed by \citet{panopoulou2017}, $f\subbroad\sim1.5\%$.

However,} the factor of two difference between the two observationally-determined exponents {--- $\fwhm^{-5}$ from \citet{panopoulou2017}, as against $\fwhm^{-2.6}$ from \citet{arzoumanian2019} ---} suggests that there are difficulties with {collating and} measuring unambiguously the widths of broader, more diffuse filaments. The flows creating real filaments are less well co-ordinated and coherent than the homogeneous, cylindrically symmetric flows of our model, and therefore the detection and characterisation of very young, forming filaments, with low density contrasts and broad profiles, is very fraught. In {particular,} most measurements of filament $\fwhm$s depend on an automated algorithm to locate and trace the spine of a filament before measuring its width. Very young, broad filaments with low density contrast, low signal-to-noise {and poor coherence} are hard to trace, basically because the second derivative of the density field is {noisy and} difficult to estimate. Such filaments are therefore likely to escape detection.

Furthermore, the threshold surface-density that we use for a filament to be an observable filament, $\Sigma\subSPINE>\Sigma\subMIN\simeq 37.2 \msun \pc^{-2}$, is about half the typical value used by \citet{arzoumanian2019} ($\Sigma\subMIN\sim73 \msun \pc^{-2}$ for a background column-density $N_{\rm H_2} = 3 \times 10^{21} \pcc$ and density contrast $0.3$). Increasing our threshold would preferentially exclude filaments with higher $\fwhm$ values, resulting in a distribution with a steeper tail, and fewer filaments with $\fwhm>0.2 \pc$.

{In contrast, the distribution of $\fwhm$s produced by the {\it subcritical} filament model (see Figure \ref{fig:histsub}) has $48\%$ of filaments with $\fwhm >0.2\pc$. Moreover, as discussed above, the shape of the overall distribution is also very different from the observed distributions reported by \citet{panopoulou2017} and \citet{arzoumanian2019}. While faint, incoherent, broad filaments are difficult to detect, it is very unlikely that the undetected fraction would be sufficiently large to transform Figure \ref{fig:histsub} into the observed distributions, since a significant proportion of these broad filaments also have high surface densities (see Figure \ref{fig:fwsd}). Again, it appears that the observations cannot be reconciled with subcritical filament formation.}

\subsection{The relationship between $\fwhm$ and $\Sigma$}

{The observed $\fwhm$s of filaments are independent of their central surface-density \citep{arzoumanian2019}, whereas isothermal cylinders in equilibrium should have a width scaling with the {thermal} Jeans length on the spine,
\begin{eqnarray}\label{EQN:lambdaJEANS}
\fwhm\;\;\propto\;\;\lambda\subJEANS&\sim&\frac{a\subO^2}{G\Sigma\subSPINE}.
\end{eqnarray}
Figure \ref{fig:fwsd} shows $\fwhm$ values plotted against the corresponding $\Sigma\subSPINE$ values, for the observable filaments used to generate the $\fwhm$ distributions plotted in Figures \ref{fig:hist} and \ref{fig:histsub}. In all cases, we find that the {\it maximum} $\fwhm$ at a given $\Sigma\subSPINE$ does decline approximately as $\Sigma\subSPINE^{-1}$. However, over the range of surface-densities covered by the data in \citet{arzoumanian2019} ($\sim 10\msun \pc^{-2}$ to $\sim 400 \msun \pc^{-2}$), there is always a range of possible $\fwhm$s between $\sim 0.03\pc$ and $\sim 0.1 \pc$. If, as we have already argued, filaments with small $\Sigma\subSPINE$ and/or large $\fwhm$ are likely to evade detection, our results are approximately consistent with the lack of correlation between $\fwhm$ and $\Sigma\subSPINE$ reported by \citet{arzoumanian2019}.}

{Unlike the ${\cal B}\subperp = 1.5$ subcritical models shown in Figure \ref{fig:fwsd}, the \citet{auddy2016} analytic model predicts a nearly-constant $\fwhm$ for all values of $\Sigma\subSPINE$. However, \citet{auddy2016} only consider inflows in the direction perpendicular to the magnetic field, and assume that the filament has time to relax to hydrostatic equilibrium in the direction parallel to the field. Our filaments are never in hydrostatic equilibrium, and so will be significantly more compressed when viewed at the point of maximum compression, and with the magnetic field close to the plane of the sky. These viewing angles account for the high $\Sigma\subSPINE$, low $\fwhm$ points in Figure \ref{fig:fwsd}, and have no equivalent in the analytic treatment.}

\section{Conclusions}

Filamentary structures observed in molecular clouds have {three} notable properties. {(i) They have $\fwhm$s of approximately $0.1 \pc$. (ii) The dominant filaments are generally aligned perpendicularly to the local magnetic field. (iii) Their $\fwhm$s appear not to be correlated with their spinal surface-densities.} We have previously argued that converging supersonic turbulent motions naturally result in a distribution of filament $\fwhm$s peaking at around, or somewhat below, $0.1 \pc$ \citep{priestley2022}.

{Here we show that introducing a perpendicular magnetic field to the model does not substantially change this conclusion, provided that filaments have supercritical mass-to-flux ratios. Although magnetised but supercritical filaments are broader than non-magnetised ones, when viewed from directions close to the magnetic field, the overall distribution of randomly oriented filament $\fwhm$s is still quite sharply peaked at values comparable to non-magnetised models.

However, if filament mass-to-flux ratios are subcritical, they have a much broader $\fwhm$ distribution, peaking at significantly larger values. This is incompatible with the extremely small number of observed filaments with $\fwhm \gtrsim 0.2 \pc$.

We conclude that our dynamical filament formation model is able to reproduce the main properties of the observed filament $\fwhm$ distribution, provided that the observed filaments are magnetically {\it supercritical}. This suggests that magnetically {\it subcritical} filaments are either very rare, or very difficult to observe (due to their low column-densities and large $\fwhm$s).}

\section*{Acknowledgements}
FDP and APW acknowledge the support of a Consolidated Grant (ST/K00926/1) from the UK Science and Technology Facilities Council (STFC). We are grateful to Gina Panopoulou for sharing observational data, and for an informative discussion on its interpretation. We thank the anonymous referee for a prompt and constructive report which helped us to improve an earlier version of the paper.

\section*{Data Availability}
The data underlying this article will be shared on request.

\bibliographystyle{mnras}
\bibliography{DynFilFormExp1}

\appendix

\section{Ambipolar diffusion during the formation of a supercritical filament}\label{sec:nimhd}

{In the initial configuration, the ambipolar drift velocity is $\upsilon\subAD\sim W\subO/t\subADO$, where $t\subADO$ is the initial ambipolar diffusion timescale. In this appendix we show that for supercritical filaments (${\cal B}\subperp<1$) $\,t\subADO$ is significantly longer than the dynamical timescale, $t\subDYNO$, and therefore non-ideal MHD effects can reasonably be ignored during the assembly of a filament.

The coupling time between the neutrals and the ions is
\begin{eqnarray}
t\subCOUPLE&\sim&\frac{1}{n_{\rm i}\left<\sigma\upsilon\right>}\;\;=\;\;\frac{1}{\nh\,x_{\rm i}(\nh)\left<\sigma\upsilon\right>}\,,\hspace{1.5cm}
\end{eqnarray}
where, $x_{\rm i}$ is given by Equation~\ref{EQN:xi}, and $\left<\sigma\upsilon\right>\sim 1.5\times 10^{-9}\,\rm{cm^3\,s^{-1}}$ is the rate coefficient for elastic collisions between ions and neutrals \citep{draine1983}. The drag acceleration on the ions is therefore
\begin{eqnarray}\label{EQN:aDRAG.01}
a\subDRAG&\sim&\frac{\upsilon\subAD}{t\subCOUPLE}\;\;\sim\;\;\frac{W\subO\,\nh\,x_{\rm i}(\nh)\,\left<\sigma\upsilon\right>}{t\subADO}.\hspace{0.8cm}
\end{eqnarray}

The magnetic acceleration on the ions is
\begin{eqnarray}\label{EQN:aMAG.01}
a\subMAG&\sim&\frac{B\subOperp^2}{4\,\pi\,\nh\,\bar{m}\subH\,W\subO},
\end{eqnarray}
where $\bar{m}\subH\sim 2.37\times 10^{-24}\,\rm{g}$ is the mass associated with each hydrogen atom when account is taken of other elements (in particular helium). Equating the two accelerations (Equations~\ref{EQN:aDRAG.01} and \ref{EQN:aMAG.01}), we obtain
\begin{eqnarray}\label{EQN:tAD.01}
t\subADO&\sim&\frac{4\,\pi\,\bar{m}\subH\,\left<\sigma\upsilon\right>\,\nh^2\,x_{\rm i}(\nh)\,W\subO^2}{B\subOperp^2}\\\label{EQN:tAD.02}
&\sim&26\,\rm{Myr}\left[\frac{W\subO}{\rm pc}\right]^{1.2}\,{\cal G}^{\,-0.6}\,{\cal B}\subperp^{\,-2}; 
\end{eqnarray}
Equation~\ref{EQN:tAD.02} is obtained from Equation~\ref{EQN:tAD.01} by substituting 
\begin{eqnarray}
n\subH\!\!&\!\!\sim\!\!&\!\!\frac{\mu\subO}{\pi W\subO^2\bar{m}\subH}\,\sim\,\frac{2\,a\subO^2\,{\cal G}}{\pi G\bar{m}\subH\,W\subO^2}\,\sim\,148\,\rm{cm^{-3}}\!\left[\!\frac{W\subO}{\rm pc}\!\right]^{-2}\!{\cal G},\hspace{0.7cm} 
\end{eqnarray}
and (from Equation~\ref{EQN:BO.02})
\begin{eqnarray}
B\subOperp\!&\!\sim\!&\!\frac{3\,\pi\,a\subO^2\,{\cal G}\,{\cal B}\subperp}{[5G]^{1/2}\,W\subO}\;\sim\;1.85\,\mu\rm{G}\left[\!\frac{W\subO}{\rm pc}\!\right]^{\,-1}{\cal G}\,{\cal B}\subperp.\hspace{0.4cm} 
\end{eqnarray}
With $W\subO =1\,\rm{pc}$, ${\cal G}=1.2$ and ${\cal B}\subperp=0.7$, Equation~\ref{EQN:tAD.02} gives $t\subADO\sim 48\,\rm{Myr}$. 

The dynamical timescale is initially of order
\begin{eqnarray}
t\subDYNO&\sim&\frac{W\subO}{{\cal M}\,a\subO}\;\;\sim\;\;5.2\,\rm{Myr}\left[\frac{W\subO}{\rm pc}\right]\,{\cal M}^{\,-1},\hspace{1.0cm}
\end{eqnarray}
and therefore, even for the slowest inflow speeds (${\cal M}\sim 2$), significantly shorter than the ambipolar diffusion timescale $t\subDYNO/t\subADO\lesssim 0.06$. 

If the cylinder contracts radially (and if we ignore longitudinal motions; see below), the magnetic field, $B\subperp$, increases in inverse proportion to the radius, $W$, i.e. $B\subperp=B\subOperp[W/W\subO]^{-1}$ and therefore the ambipolar diffusion timescale decreases as $t\subAD\simeq t\subADO[W/W\subO]^{1.2}$. The dynamical timescale scales as $t\subDYN\simeq t\subDYNO[W/W\subO]$, and possibly a little faster due to inward gravitational acceleration, at least until the inflowing material hits the accretion shock \citep[see][]{priestley2022}. Consequently the ratio of the timescales, $t\subDYN/t\subAD$ is approximately constant, up to the accretion shock (which is the locus that essentially defines the filament $\fwhm$).

If the cylinder also contracts longitudinally, the number-density, $n$, increases as $n\sim n_{\rm o}[L/L\subO]^{-1}$, and the transverse magnetic field, $B\subperp$ increases as $B\subperp\simeq B\subOperp[L/L\subO]^{-1}$, so $t\subAD$ decreases slowly, as $t\subAD\simeq t\subADO[L/L\subO]^{1/2}$. This is a small effect in our simulations, since by design we have explored a setup in which longitudinal contraction is small.

We conclude that ambipolar diffusion can be ignored during the filament formation phase, if the filament is supercritical. Once a supercritical filament has formed, with radius $W\ll W\subO$, it will collapse and fragment on a timescale 
\begin{eqnarray}
t\subCOLL\!&\!\!\sim\!\!&\!\frac{[W\!/W\subO]}{[\pi G\rho\subO]^{1/2}[1-{\cal B}\subperp^{\,2}]^{1/2}}\;\sim\;\frac{3.4\,\rm{Myr}\,[W\!/W\subO]}{{\cal G}^{1/2}[1-{\cal B}\subperp^{\,2}]^{1/2}}.\hspace{0.7cm}
\end{eqnarray}
With $W\lesssim 0.05\,\rm{pc}$, $W\subO = 1\,\rm{pc}$ and ${\cal G}>1$, the collapse timescale is very short compared with the ambipolar diffusion timescale, unless the filament is only marginally supercritical. We conclude that ambipolar diffusion can only play an important role in the dynamics after the filament has fragmented into cores.

We reiterate that the arguments presented here pertain only to the case of a {\it supercritical} filament. For a {\it subcritical} filament (${\cal B}\subperp>1$), the ambipolar diffusion time-scale is shorter (see Equation~\ref{EQN:tAD.02}), and notionally (i.e. for the idealised case of an infinitely long, cylindrically symmetric filament) the ideal-MHD collapse timescale is infinite, because the filament should just wobble around until it reaches a static equilibrium. In this case, ambipolar diffusion will be important, but, as we have shown in Sections \ref{SEC:SubFils}, \ref{SEC:PDFPeak} and \ref{SEC:FatFils}, the timescale for this is implausibly long, and the resulting distribution of filament $\fwhm$s is incompatible with the observations.}

\begin{figure}
  \centering
  \includegraphics[width=0.45\textwidth]{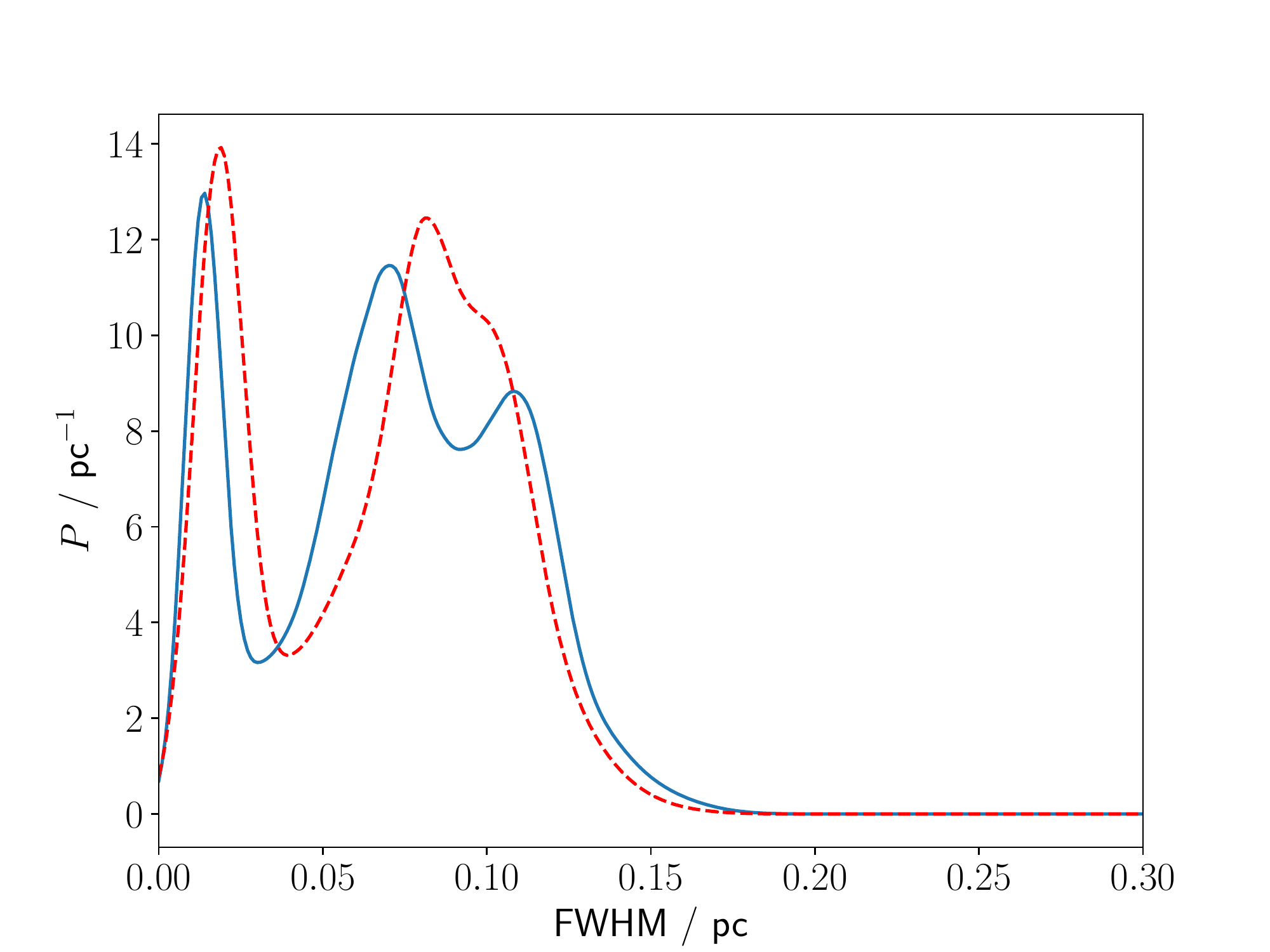}\quad
  \caption{The distributions of filament $\fwhm$ for the supercriticial model with ${\cal M}=3$ and ${\cal B}\subperp=0.5$, simulated with ideal MHD (full blue curve) and with non-ideal MHD (dashed red curve), and viewed from a flat distribution of viewing angles.}
  \label{FIG:NonIdealSupercritical}
\end{figure}

\section{A supercritical filament evolved with non-ideal MHD}\label{APP:NonIdeal}

{In order to justify the neglect of non-ideal effects in the supercritical models, we have rerun the specific combination ${\cal M}=3$ and ${\cal B}\subperp=0.5$ using non-ideal MHD. Figure \ref{FIG:NonIdealSupercritical} shows the resulting distributions of $\fwhm$. The dashed red curve on Figure \ref{FIG:NonIdealSupercritical} shows the distribution when the model is evolved with non-ideal MHD. The solid blue curve on Figure \ref{FIG:NonIdealSupercritical} shows the distribution when the model is evolved with ideal MHD; this is the same distribution as presented on both panels of Figure \ref{fig:pdf} (blue dashed curve on the left panel, blue solid curve on the right panel). Evidently the differences are very small, and will become even smaller when the distributions obtained with different ${\cal M}$ values are averaged over a flat distribution. We conclude that non-ideal effects can safely be neglected in the supercritical models.}

\bsp	
\label{lastpage}
\end{document}